\documentclass[journal]{IEEEtran}

\usepackage{amsmath}
\usepackage{multicol}
\usepackage{multirow}
\newcommand{\tabincell}[2]{\begin{tabular}{@{}#1@{}}#2\end{tabular}}
\usepackage{graphicx}
\usepackage{epstopdf}
\usepackage{extarrows}
\usepackage{float}
\usepackage[latin1]{inputenc}
\allowdisplaybreaks[4]

\begin{document}

\title{Bi-Fuzzy Discrete Event Systems and Their Supervisory Control Theory}

\author{Weilin Deng
        and Daowen Qiu$^{\star}$ 
\thanks{Weilin Deng is  with the Department of Computer Science, Sun Yat-sen University,
Guangzhou, 510006, China, and with the Department of Computer Engineering, Guangdong Industry Technical College, Guangzhou, 510300, China.  (e-mail:williamten@163.com)}
\thanks{Daowen Qiu (Corresponding author) is with the Department of Computer science, Sun Yat-sen
 University, Guangzhou, 510006, China. (e-mail: issqdw@mail.sysu.edu.cn)}}

\maketitle

\begin{abstract}
It is well known that type-1 fuzzy sets (T1 FSs) have limited capabilities to handle some data uncertainties directly, and type-2 fuzzy sets (T2 FSs) can cover the shortcoming of T1 FSs to a certain extent.
 Fuzzy discrete event systems (FDESs) were proposed  based on T1 FSs theory. Hence, FDES may not be a satisfactory model to characterize some
 high-uncertainty systems.
 In this paper, we propose a new model, called as bi-fuzzy discrete event systems (BFDESs), by combining classical DESs theory and T2 FSs theory.
  Then, we consider the supervisory control problem of BFDESs. The bi-fuzzy controllability theorem and nonblocking bi-fuzzy controllability theorem are demonstrated.
   Also,  an algorithm for checking the bi-fuzzy controllability condition is presented.
  In addition, two controllable approximations to an uncontrollable language are investigated in detail.
   An illustrative example is provided to show the applicability and the advantages of BFDESs model.
\end{abstract}


\IEEEpeerreviewmaketitle

\newtheorem{definition}{Definition}
\newtheorem{theorem}{Theorem}
\newtheorem{lemma}{Lemma}
\newtheorem{corollary}{Corollary}
\newtheorem{remark}{Remark}
\newtheorem{algorithm}{Algorithm}
\newtheorem{example}{Example}
\newtheorem{proposition}{Proposition}

\section{Introduction}

\IEEEPARstart{D}{iscrete} event systems (DESs) are dynamic event-driven systems with discrete states.
The supervisory control problem of DESs has received much attention in the last twenty
years [1-3]. 
In supervisory
 control theory, an uncontrolled system (usually called as a plant) is modeled as a finite automaton. A desired system behavior (usually called as a specification) is given by a formal language.
However, in real-world situation, there are a large number of systems associated with  vagueness, impreciseness, and subjectivity. The behaviors of such systems cannot be captured by finite automata or formal languages.
 In order to characterize such systems, Lin and Ying \cite{Feng-1}
first proposed the fuzzy discrete event systems (FDESs).
After that, Qiu and Liu [5-7], and Cao and Ying [8-9], respectively, developed the supervisory control theory of FDESs. Recently, Jayasiri established modular and hierarchical supervisory control theory of FDESs in \cite{Jayasiri-1}, and  generalized  the decentralized control theory of FDESs in \cite{g-de}.  Moreover, FDESs have been applied to practical problems in many areas, such as decision making \cite{decision-making}, disease treatment supporting [13-15], robotic control [17-18], and traffic management \cite{traffic}, etc.
\par

 It is necessary to point out that the FDESs models mentioned above are based on T1 FSs theory, which was first proposed by Zadeh  \cite{zadeh-t1fs}.
 Although T1 FSs have been successfully used in many fields (see \cite{fuzzy-book} and its references),  T1 FSs  have limited capabilities to directly handle some linguistic and data uncertainties because the membership functions they use are certain \cite{Mendel-3}.
For example, suppose that $X = \{Lucy, Maria  , Anne\}$ is a set of girls, and $\tilde{B}$ is a T1 FS of beautiful girls in X and $\tilde{B}$ is captured by the following membership function:
\begin{equation}
        \tilde{B} = \frac{\mu_{\tilde{B}}(Lucy)}{Lucy} + \frac{\mu_{\tilde{B}}(Maria)}{Maria}+\frac{\mu_{\tilde{B}}(Anne)}{Anne}.
\end{equation}
Here the membership degrees $\mu_{\tilde{B}}(Lucy)$, $\mu_{\tilde{B}}(Maria)$ and $ \mu_{\tilde{B}}(Anne)$ are 
crisp numbers in interval $[0,1]$. On the other hand, the word ``beautiful" itself is uncertain, for words can mean different things to different people. That is, $\mu_{\tilde{B}}(Lucy)$, as well as $\mu_{\tilde{B}}(Maria)$ and $ \mu_{\tilde{B}}(Anne)$, may be specified with different numbers by different people.
Since the FDESs are formulated based on T1 FSs, FDESs  cannot directly model the uncertainties of some physical systems. Therefore, FDESs might not be so satisfactory models for some high-uncertainty systems.\par
To make up for the significant drawback of T1 FSs, T2 FSs were first proposed by Zadeh  \cite{zadeh-1}. The membership degrees of a T2 FS are T1 FSs in  [0,1] rather than crisp numbers in [0,1]. Hence, T2 FSs can be used to handle uncertainties directly  in a better way because they provide us with more parameters in modeling. In recent decades, T2 FSs have been widely investigated (see [24-25] and their references). Notably, R. Sepu\'{u}lveda et al. \cite{experimental-study} made an experimental study of T1 and T2 fuzzy logic systems, which shows that the best results are obtained by using T2 fuzzy systems. In addition, Mendel  \cite{Quantitative-Comparison} made a quantitative comparison of T2 and T1 fuzzy systems. The theoretical results in \cite{Quantitative-Comparison} suggest that the higher uncertainty a physical system has, the more a T2 FLS outperforms a T1 FLS.
Recently, Du et al. \cite{knoledge} generalized FDESs to Extended FDESs (EFDESs) based on T2 FSs by allowing all the elements in fuzzy state vectors and fuzzy event transition matrices to be fuzzy numbers.
In \cite{knoledge}, the max-min and max-product operations were defined by using the Zadeh's Extension Principle.
\par
In this paper, we also present a generalized FDESs model, called as Bi-Fuzzy DESs (BFDESs), based on T2 FSs. Different from Du's model \cite{knoledge},  BFDESs allow all the elements in fuzzy state vectors and fuzzy event transition matrices to be normal convex T1 FSs,
   and the operations of BFDESs are derived from Mizumoto's \cite{Mizumoto-1} and Mendel's methods \cite{Mendel-1}, \cite{Mendel-2}. The main purpose of the paper is to introduce the BFDESs model and establish their supervisory control theory.\par

    The main contributions of the paper are as follows.\par
\begin{enumerate}
\item
 In Section \uppercase\expandafter{\romannumeral3}, the BFDESs model are introduced by  combining classical DESs theory and T2 FSs theory.
    Then, the fundamental properties of BFDESs are discussed, and the parallel composition operation of BFDESs is formulated.

\item
In Section \uppercase\expandafter{\romannumeral4}, the supervisory control problems of BFDESs are studied. The controllability theorem and nonblocking controllability theorem of BFDESs are demonstrated, and thus the bi-fuzzy controllability condition is obtained. Furthermore, an algorithm for checking the condition is introduced.

\item
    In Section \uppercase\expandafter{\romannumeral5}, in order to demonstrate the applicability of the supervisory control theory of BFDESs, a traffic signal control approach based on BFDESs model is proposed. Also, another approach based on FDESs model can be directly constructed via reducing the BFDESs model to FDESs model.
    Then, a simulation experiment is carried out and the results show that BFDESs model have well  advantages over FDESs model in general.
    \item
In  Appendix A,
the supremal controllable bi-fuzzy sublanguage and the infimal prefix-closed controllable bi-fuzzy superlanguage are defined and investigated in detail. The two controllable languages are demonstrated to be the best approximations to an uncontrollable bi-fuzzy language.  Thus, they could  act as the alternative schemes if the given specifications cannot be achieved by supervisory control.
\end{enumerate}

    \par
    Besides the sections mentioned before, Section \uppercase\expandafter{\romannumeral2} provides the preliminary  knowledge.
 Section \uppercase\expandafter{\romannumeral6} summarizes the main results obtained. \par

\section{Preliminaries}
  In this section, we would briefly review the necessary knowledge about T2 FSs.  For  more details,
    we can refer to [22], [29-31]. \par
 T2 FSs have
membership degrees that are T1 FSs in $[0,1]$. Hence, T2 FSs are usually called as \emph{bi-fuzzy sets}. Formally, the definition of  T2 FSs is presented as follows.

\begin{definition}
  A \emph{T2 FS} of a set $X$, denoted by $\hat{A}$, can be expressed as
  \begin{equation}
    \hat{A} = \int_{x\in X}\int_{u  \in J_x} [ \mu_{\hat{A}}(x,u)/u ]/x,
  \end{equation}
  where $J_x \subseteq [0,1] , \ \mu_{\hat{A}}(x,u) \in [0,1]$.
  Here,   $u$ is the \emph{primary membership }of $x$ in $\hat{A}$.  $\mu_{\hat{A}}(x,u)$ is the \emph{secondary membership} of $x$ in $\hat{A}$ with respect to the primary membership $u$.
  The \emph{fuzzy  degree} of $x \in X$ in $\hat{A}$ is defined as
  \begin{equation}
       \mu_{ \hat{A}}(x) =  \int_{u\in J_x} \mu_{\hat{A}}(x,u) /u.  
  \end{equation}
  Thus, a fuzzy  degree can be regarded as a T1 FS in $J_x$.
  Then $\hat{A}$ can also be expressed as
    \begin{equation}
      \hat{A} = \int_{x\in X} \mu_{ \hat{A}}(x)  /x.
    \end{equation}\par
  The $\int$ in the above three equations should be replaced by $\sum$ for discrete cases.
\end{definition}
\par
In this paper, we only consider \emph{normal convex fuzzy degrees} (\emph{NCFD}). That is, the $ \mu_{ \hat{A}}(x)$ is \emph{normal}, i.e.,
$\max_{u\in J_x} \mu_{\hat{A}}(x,u) = 1 $, and $\mu_{ \hat{A}}(x)$ is \emph{convex}, i.e., $\mu_{\hat{A}}(x,u_j) \geq \min\{
\mu_{\hat{A}}(x,u_i),\mu_{\hat{A}}(x,u_k)\}$ holds for $\forall u_i\leq u_j \leq u_k$, $u_i,u_j,u_k \in J_x $.
\par

Consider two T2 FSs, $\hat{A} = \int_{x\in X}\mu_{ \hat{A}}(x)/x$ and $\hat{B} = \int_{x\in X}\mu_{ \hat{B}}(x)/x$, where $\mu_{\hat{A}}(x) = \int_u \mu_{\hat{A}}(x,u)$ $/u $ and $\mu_{\hat{B}}(x) = \int_w \mu_{\hat{B}}(x,w) /w $.  Mizumoto \cite{Mizumoto-1}  and Mendel \cite{Mendel-1}  defined the union ($\cup$), intersection ($\cap$) and complements ($\bar{\ \  }$) operations of T2 FSs as follows:
\begin{multline*}
    \hat{A} \cup \hat{B}  = \int_{x \in X} \mu_{\hat{A}}(x) \sqcup \mu_{\hat{B}}(x)/x,\  \text{where}\ \\
     \mu_{\hat{A}}(x) \sqcup \mu_{\hat{B}}(x) =  \int_u\int_w(\mu_{\hat{A}}(x,u)  \wedge \mu_{\hat{B}}(x,w) )/(u \vee w),
\end{multline*}
\begin{multline*}
    \hat{A} \cap \hat{B}= \int_{x \in X} \mu_{\hat{A}}(x) \sqcap \mu_{\hat{B}}(x)/x, \ \text{where} \\
    \mu_{\hat{A}}(x) \sqcap \mu_{\hat{B}}(x) = \int_u\int_w(\mu_{\hat{A}}(x,u)  \wedge \mu_{\hat{B}}(x,w) )/(u \wedge w),
\end{multline*}
\begin{equation*}
   \bar{\hat{A}}  = \int_{x \in X}\mu_{\bar{\hat{A}}}(x)/x, \ \text{where}\ \mu_{\bar{\hat{A}}}(x) = \int_u \mu_{\hat{A}}(x,u)/(1-u).
\end{equation*}
\par
Here  $\vee$ and $\wedge$ denote $\max$ operation and $\min$ operation, respectively.
 $\sqcap$ and $\sqcup$  defined in fuzzy  degrees are called as join and meet, respectively.
\par

According to the join operation introduced above, the inclusion relation $``\sqsubseteq"$ between two fuzzy  degrees is defined as follows.
\begin{equation}
  \mu_{\hat{A}}(x) \sqsubseteq \mu_{\hat{B}}(x) \xLongrightarrow{define}  \mu_{\hat{A}}(x) \sqcap \mu_{\hat{B}}(x) = \mu_{\hat{A}}(x).
\end{equation}
\par
Furthermore, the inclusion relation $``\subseteq"$ between two T2 FSs is defined as follows.
\begin{equation}
  \hat{A}\subseteq \hat{B} \xLongrightarrow{define} \mu_{\hat{A}}(x) \sqsubseteq \mu_{\hat{B}}(x),\ \  \forall x \in X .
\end{equation}

\par
 Table  \uppercase\expandafter{\romannumeral1}, obtained from \cite{Mizumoto-1} and \cite{Mendel-1}, shows the main properties of \emph{NCFD}.
\begin{table}
\begin{center}
\caption{Properties of \emph{NCFD} ($\mu_{1}, \mu_{2}, \mu_{3} \in$ \emph{NCFD}). }
\begin{tabular}{c c c}
\hline
\textbf{ laws} & \textbf{description }  \\
\hline
Reflexive & $\mu_{1}  \sqsubseteq \mu_{1}  $   \\
\hline

Antisymmetric & $\mu_{1}  \sqsubseteq \mu_{2} \  ,\ \mu_{2}  \sqsubseteq   \mu_{1}  \Rightarrow \mu_{1}  = \mu_{2} $    \\
\hline
Idempotent & $\mu_{1}  \sqcap \mu_{1}  = \mu_{1} \ ;\ \mu_{1}  \sqcup \mu_{1}  = \mu_{1} $   \\
\hline

 Transitive & $\mu_{1}  \sqsubseteq \mu_{2} \ ,\ \mu_{2}  \sqsubseteq \mu_{3}  \Rightarrow \mu_{1}  \sqsubseteq \mu_{3} $    \\
\hline

Identity & \tabincell{c}{$\mu_{1}  \sqcup \frac{1}{0} = \mu_{1}  \ ;\ \mu_{1}  \sqcap \frac{1}{0} = \frac{1}{0}\ $\\ $ \mu_{1}  \sqcup \frac{1}{1} = \frac{1}{1} \ ;\ \mu_{1}  \sqcap \frac{1}{1} = \mu_{1} $} \\
\hline

Commutative & $\mu_{1}  \sqcap \mu_{2}  = \mu_{2}  \sqcap \mu_{1};\mu_{1}  \sqcup \mu_{2}  = \mu_{2}  \sqcup \mu_{1} $   \\
\hline

Absorption & \tabincell{c}{$\mu_{1}  \sqcap (\mu_{1}  \sqcup \mu_{2} ) = \mu_{1}$ \\  $   \mu_{1}  \sqcup (\mu_{1}  \sqcap \mu_{2} ) = \mu_{1} $}   \\
\hline

Associative & \tabincell{c}{$(\mu_{1}  \sqcap \mu_{2} ) \sqcap \mu_{3}  = \mu_{1}  \sqcap (\mu_{2}  \sqcap \mu_{3} ) $ \\ $ (\mu_{1}  \sqcup \mu_{2} ) \sqcup \mu_{3}  = \mu_{1}  \sqcup (\mu_{2}  \sqcup \mu_{3} )$ } \\
\hline

Distributive & \tabincell{c}{$\mu_{1}  \sqcup (\mu_{2}  \sqcap \mu_{3})=(\mu_{1}  \sqcup \mu_{2}) \sqcap (\mu_{1}  \sqcup \mu_{3})$  \\
$\mu_{1}  \sqcap (\mu_{2}  \sqcup \mu_{3})=(\mu_{1}  \sqcap \mu_{2} ) \sqcup (\mu_{1}  \sqcap \mu_{3})$} \\
\hline

\end{tabular}
\end{center}
\end{table}
\par
As the generalized T1 fuzzy relation, a T2 fuzzy relation is defined as follows.
\begin{definition}
  Let $X$ and $Y$ be two universes of discourse. Then
  \begin{equation}
    \hat{R} = \{ [(x,y),\mu_{\hat{R}}(x,y)]\ |\ (x,y) \in X \times Y  \}
  \end{equation}
  is a binary T2 fuzzy relation in the product space $X \times Y$. Here, $\mu_{\hat{R}}(x,y) \in$ \emph{NCFD} denotes the fuzzy degree of $(x,y)$ belonging to  $\hat{R}$.
  Let $NCFD^{X\times Y}$ denote the set of all T2 fuzzy relations in the product space $X \times Y$. Then  $\hat{R}\in NCFD^{X\times Y}$ with $|X|=m$ and $|Y|=n$ can also be expressed as an $m*n$ matrix in which the elements belong to \emph{NCFD}.
\end{definition}
\par
Suppose that $\hat{R} \in NCFD^{X\times Y}$ and $\hat{S} \in NCFD^{Y \times Z}$  with $|X| = m$, $|Y| = n$ and $|Z| = k$. Then the composition of $\hat{R}$ and $\hat{S}$ is denoted by the $m*k$ matrix $\hat{R}\hat{\odot} \hat{S}$, in which the elements are obtained by the following meet-join operation:
\begin{equation}
  \hat{R} \hat{\odot} \hat{S}(x,z) = \bigsqcup_{y \in Y} [ \hat{R}(x,y) \sqcap \hat{S} (y,z)],\ \  x \in X, z \in Z.
\end{equation}

\section{Bi-Fuzzy Discrete Event Systems}
As mentioned in Section \uppercase\expandafter{\romannumeral1}, FDES might not be  a satisfactory model to characterize  high-uncertainty systems, so, in this section, we introduce a new model and investigate some of the main properties of this model.
\par
 Since T2 FSs could be called as bi-fuzzy sets, our model based on T2 FSs is named as bi-fuzzy DESs. Formally, we have the following notion.
 \begin{definition}
        A \emph{bi-fuzzy DES }(BFDES) is modeled as a bi-fuzzy automaton, which is a five-tuple:
        \[
        \hat{G}=\{\hat{X}, \hat{\Sigma}, \hat{\delta}, \hat{x}_{0}, \hat{x}_{m}\}.
        \]
        Here $\hat{X}$ is a set of bi-fuzzy states over a crisp state set $X$ with $|X| = n$. A bi-fuzzy state $\hat{x}\in \hat{X}$ is denoted by  a row vector $\{\tilde{x}_1,\tilde{x}_2,\ldots ,\tilde{x}_n\}$, where $\tilde{x}_i \in NCFD$ represents the fuzzy degree of the system  being at the crisp state $x_i$.
                $\hat{\Sigma}$ is a set of bi-fuzzy events. Any $\hat{\sigma}\in \hat{\Sigma}$ is denoted by a matrix $\hat{\sigma}=[\tilde{a}_{ij}]_{n*n}$ with $\tilde{a}_{ij}\in NCFD $.  $\tilde{a}_{ij}$ denotes the fuzzy transition degree from  state $x_i$ to $x_j$ when event $\hat{\sigma}$ occurs.
        $\delta:\hat{X}\times\hat{\Sigma}\rightarrow \hat{X}$ is a transition function, which is defined by $\hat{\delta}(\hat{x},  \hat{\sigma})=\hat{x} \hat{\odot} \hat{\sigma}$ for $\hat{x}\in \hat{X}$ and $\hat{\sigma} \in \hat{\Sigma}$.
         ``$\hat{\odot}$" denotes the meet-join operation defined in Equation (8).
         $\hat{x}_{0} = [\tilde{x}_{0,1},\tilde{x}_{0,2},\ldots,\tilde{x}_{0,n}]$ is a bi-fuzzy initial state, where $\tilde{x}_{0,i} \in NCFD $ is the fuzzy degree of the crisp state $x_{i}$ belonging to initial states.
        $\hat{x}_{m} = [\tilde{x}_{m,1},\tilde{x}_{m,2},\ldots,\tilde{x}_{m,n}]$ is the bi-fuzzy final state, where $\tilde{x}_{m,i} \in NCFD $ is the fuzzy degree of the crisp state $x_{i}$ belonging to final states.

 \end{definition}
 \begin{remark}
    Du, Ying and Lin \cite{knoledge} also generalized FDESs model to extended FDESs (EFDESs) model based on T2 FSs.
   However, our model has two main different points from  Du's model: (1) BFDESs allow all the elements in fuzzy state vectors and fuzzy event transition matrices to be normal convex T1 FSs rather than fuzzy numbers. Hence, BFDESs are more general. (2) BFDESs  use the  meet-join operation rather than max-min or max-product operation to characterize the event-driven evolutions of bi-fuzzy states.
 \end{remark}
 \par
  BFDESs can  handle uncertainties directly in a better way than FDESs because they provide us with more parameters in modeling event-driven systems. A numerical example concerning the modeling of medical treatments is presented as follows.
 \begin{example}
 For a newly-found disease, physicians  cannot give a exact score to evaluate the therapeutic effect of a treatment regimen due to their limited knowledge about the disease. Assume four physicians (that is, 1, 2, 3, and 4)  give their evaluations to  two treatment regimens (that is, A and B) in Table \uppercase\expandafter{\romannumeral2}.

\begin{table}[htp]
\begin{center}
\caption{scores of the therapeutic effects of Treatment Regimens. }
\begin{tabular}{c c c c c c c}
\hline
\textbf{ Regimens} & \textbf{ 1 } & \textbf{ 2 }& \textbf{ 3 }& \textbf{ 4 }&  \\
\hline
 A ($\tilde{\sigma}_{1}$) & $[0.6,0.8] $   & $[0.5,0.9] $  & $[0.5,0.7] $  & $[0.6,0.8] $  \\
\hline
 B ($\tilde{\sigma}_{2}$) & $[0.5,0.8] $   & $[0.5,0.7] $  & $[0.6,0.9] $  & $[0.5,0.9] $  \\
\hline

\end{tabular}

\end{center}
\end{table}
\begin{figure}
\centering
\includegraphics[width=0.4\textwidth]{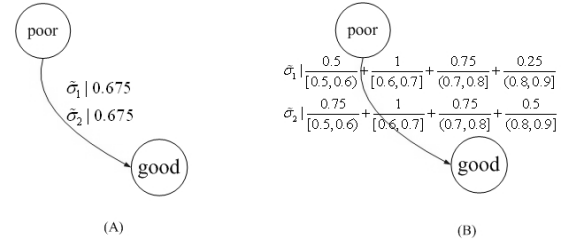}
\caption{\label{fig1} (A) the FDES model of Example 1. (B) the BFDES  model of Example 1.  The $\tilde{\sigma}_{1}$ and $\tilde{\sigma}_{2}$ denote the regimen A and regimen B, respectively. }
\end{figure}
   \par
  Assume the scores given by the four physicians are of equal importance. Then, the transition degrees from ``poor" to ``good" driven by the event $\tilde{\sigma}_{1}$  and $\tilde{\sigma}_{2}$  are obtained as follows.
   \begin{multline*}
   \tilde{\sigma}_{1,12} =   ((0.6 + 0.8)/2 + (0.5 + 0.9)/2  + \\
                              (0.5 + 0.7)/2 + (0.6 + 0.8)/2 ) / 4  = 0.675;
   \end{multline*}
   \begin{multline*}
   \tilde{\sigma}_{2,12} =   ((0.5 + 0.8)/2 + (0.5 + 0.7)/2  + \\
                              (0.6 + 0.9)/2 + (0.5 + 0.9)/2 ) / 4  = 0.675.
   \end{multline*}
    Then, we model the situation to an FDES (see Fig. 1-(A)). Obviously, the FDES loses the uncertainty of the physical system due to the premature defuzzification to original fuzzy data.
 \par
   On the other hand, we also model the situation to a BFDES (see Fig. 1-(B)).
   Let us first consider the fuzzy transition degree driven by $\tilde{\sigma}_{1}$.
   Since $[0.5,0.6)$ is contained in the evaluations given by two physicians (the physician 2 and the physician 3), we specify $2/4 = 0.5$ as the secondary
membership with respect to $[0.5,0.6)$. Similarly, we specify $4/4 = 1$ to $[0.6 , 0.7]$, $3/4 = 0.75$ to $(0.7 , 0.8]$, and  $1/4 = 0.25$ to $(0.8 , 0.9]$.
   Therefore, the fuzzy transition degree from ``poor" to ``good" driven by $\tilde{\sigma}_{1}$ is
   \[
   \tilde{\sigma}_{1,12} = \frac{0.5}{[0.5,0.6)} +  \frac{1}{[0.6,0.7]} +  \frac{0.75}{(0.7,0.8]} +  \frac{0.25}{(0.8,0.9]}.
   \]
   Similarly, the fuzzy transition degree from ``poor" to ``good" driven by $\tilde{\sigma}_{2}$ could be obtained, which is
   \[
    \tilde{\sigma}_{2,12} =  \frac{0.75}{[0.5,0.6)} +  \frac{1}{[0.6,0.7]} +  \frac{0.75}{(0.7,0.8]} +  \frac{0.5}{(0.8,0.9]}.
   \]
   According to the fuzzy quantities ranking method in \cite{yager} (for more methods, see \cite{wang-1}, \cite{wang-2}), we have $\tilde{\sigma}_{2,12} \succ \tilde{\sigma}_{1,12}$ in the sense of fuzzy theory.
   \par
   Different from the FDES model, the BFDES model well captures the uncertainty of the original physical system, and hence the BFDES  can distinguish between the therapeutic effects of the two regimens. Therefore, BFDESs  might be more precise than FDESs for some cases.
   \par
   We would present another example  in Section \uppercase\expandafter{\romannumeral5} to further demonstrate the fact that sometimes it is much better to process fuzzy data directly than to defuzzify them too early.
\end{example}
\par
  \emph{Bi-fuzzy languages }are thought of  as the behavior-characterizations of BFDESs. Several concerning notions are introduced as follows.
\begin{definition}
 \emph{Bi-fuzzy languages} generated and marked by the BFDES $\hat{G} = \{\hat{X}, \hat{\Sigma}, \hat{\delta}, \hat{x}_{0}, \hat{x}_{m}\}$ with $|X|=n$, denoted by $L_{\hat{G}}$ and $L_{\hat{G}, m}$, respectively, are defined as two functions from $\hat{\Sigma}^{*}$ to $NCFD$ as follows: $ L_{\hat{G}}(\epsilon) =  L_{\hat{G},m}(\epsilon) = \frac{1}{1}$, and for
 $\forall \hat{s}=\hat{\sigma}_{1}\hat{\sigma}_{2}\ldots\hat{\sigma}_{k}\in \hat{\Sigma}^{*}$ with $k>0$,

    \begin{equation}
      L_{\hat{G}}(\hat{s})=\hat{x}_{0}\hat{\odot}\hat{\sigma}_{1}\hat{\odot}\ldots\hat{\odot}\hat{\sigma}_{k}\hat{\odot} \hat{A}_{n}^{T},
    \end{equation}
    \begin{equation}
      L_{\hat{G}, m}(\hat{s})=\hat{x}_{0}\hat{\odot}\hat{\sigma}_{1}\hat{\odot}\ldots\hat{\odot}\hat{\sigma}_{k}\hat{\odot} \hat{x}_{m}^{T}.
    \end{equation}
    Here T is the transpose operation, and the $\hat{A}_{n}=[ \underbrace{\frac{1}{1},\frac{1}{1}, \ldots, \frac{1}{1}}_{n}]$.
\end{definition}

\begin{definition}
  Bi-fuzzy language $ L_{\hat{H}} $ is called as a \emph{sublanguage} of bi-fuzzy language $L_{\hat{G}} $ if  $L_{\hat{H}}(\hat{s}) \sqsubseteq L_{\hat{G}}(\hat{s}),$  $ \forall \hat{s} \in \hat{\Sigma^{*}}$.
\end{definition}

     \begin{definition}
     The intersection ($\cap$), union ($\cup$), and connection ($\cdot$) operations of bi-fuzzy languages are defined as these functions from  $\hat{\Sigma^{*}}$ to \emph{NCFD} as follows:  $\forall \hat{s} \in \hat{\Sigma^{*}}$,
       \begin{equation}
         (L_{\hat{G}} \cap L_{\hat{H}}) (\hat{s}) = L_{\hat{G}}(\hat{s}) \sqcap L_{\hat{H}}(\hat{s}),
       \end{equation}
       \begin{equation}
         (L_{\hat{G}} \cup L_{\hat{H}}) (\hat{s}) = L_{\hat{G}}(\hat{s}) \sqcup L_{\hat{H}}(\hat{s}),
        \end{equation}
       \begin{equation}
        (L_{\hat{G}} \cdot L_{\hat{H}}) (\hat{s}) = \bigsqcup_{\hat{s}= (\hat{u} \cdot \hat{v})} [ L_{\hat{G}}(\hat{u}) \sqcap L_{\hat{H}}(\hat{v}) ].
       \end{equation}
     \end{definition}

\par
        Before investigating the properties of bi-fuzzy languages, it is necessary to study further the properties of $NCFD$ based on the results in Table \uppercase\expandafter{\romannumeral1}.
    \begin{proposition}
     Suppose $\mu_i \in NCFD, i \in {1,2,3,4}.$ Then \par
     \begin{enumerate}
     \item $\mu_1 \sqcap \mu_2 \sqsubseteq \mu_1 $.\par
     \item $\mu_1 \sqsubseteq \mu_1 \sqcup \mu_2 $.\par
     \item $\mu_1 \sqsubseteq \mu_2 , \  \mu_3  \sqsubseteq \mu_4 \Rightarrow \mu_1 \sqcap \mu_3 \sqsubseteq  \mu_2 \sqcap \mu_4 $.\par
     \item $\mu_1 \sqsubseteq \mu_2 , \  \mu_3  \sqsubseteq \mu_4 \Rightarrow \mu_1 \sqcup \mu_3 \sqsubseteq  \mu_2 \sqcup \mu_4 $.\par
     \item $\mu_1 \sqsubseteq \mu_3 , \  \mu_2  \sqsubseteq \mu_3 \Rightarrow \mu_1 \sqcup \mu_2 \sqsubseteq  \mu_3$.\par
     \item $\mu_1 \sqsubseteq \mu_2 , \  \mu_1  \sqsubseteq \mu_3 \Rightarrow \mu_1 \sqsubseteq \mu_2 \sqcap \mu_3$.\par
     \end{enumerate}
    \end{proposition}

    \begin{IEEEproof}
      See Appendix B.
    \end{IEEEproof}

\par
The following property of bi-fuzzy languages plays an important role in the sequel.
    \begin{proposition}
        For any $\hat{s}\in\hat{\Sigma}^{*}$ and any $\hat{\sigma}\in\hat{\Sigma}$,
        \begin{equation}
          L_{\hat{G}, m}(\hat{s}\hat{\sigma})\sqsubseteq L_{\hat{G}}(\hat{s}\hat{\sigma})\sqsubseteq L_{\hat{G}}(\hat{s}).
        \end{equation}
    \end{proposition}
    \begin{IEEEproof}
      See Appendix B.
    \end{IEEEproof}

    \par
The following proposition concerning bi-fuzzy languages is  derived from Proposition 1.
    \begin{proposition}
     Suppose $\hat{L}_i \in (\hat{\Sigma}^{*})^{NCFD} , i \in {1,2,3,4},$ are  bi-fuzzy languages over the common bi-fuzzy event $\hat{\Sigma}$. Then \par
     \begin{enumerate}
     \item $\hat{L}_1 \cap \hat{L}_2 \subseteq \hat{L}_1 $.\par
     \item $\hat{L}_1 \subseteq \hat{L}_1 \cup \hat{L}_2 $.\par
     \item $\hat{L}_1 \subseteq \hat{L}_2 , \ \hat{L}_3  \subseteq \hat{L}_4 \Rightarrow \hat{L}_1 \cap \hat{L}_3 \subseteq  \hat{L}_2 \cap \hat{L}_4 $.\par
     \item $\hat{L}_1 \subseteq \hat{L}_2 , \ \hat{L}_3  \subseteq \hat{L}_4 \Rightarrow \hat{L}_1 \cup \hat{L}_3 \subseteq  \hat{L}_2 \cup \hat{L}_4 $.\par
     \item $\hat{L}_1 \subseteq \hat{L}_3 ,\  \hat{L}_2  \subseteq \hat{L}_3 \Rightarrow \hat{L}_1 \cup \hat{L}_2 \subseteq  \hat{L}_3$.\par
     \item $\hat{L}_1 \subseteq \hat{L}_2 ,\  \hat{L}_1  \subseteq \hat{L}_3 \Rightarrow \hat{L}_1 \subseteq \hat{L}_2 \cap \hat{L}_3$.\par
     \end{enumerate}
    \end{proposition}

    \begin{IEEEproof}
     See Appendix B.
    \end{IEEEproof}
\par
Parallel composition is an important operation over bi-fuzzy automata. It characterizes how the bi-fuzzy systems combine with each other by synchronously executing the common events.
    For given $\hat{G}_{i}=\{\hat{X}_{i}, \hat{\Sigma}_{i}, \hat{\delta}_{i}, \hat{x}_{0i}, \hat{x}_{mi}\}, i\in\{1, 2\}$, we formulate the parallel composition of bi-fuzzy automata in terms of the following fashion:
    \begin{equation}
    \hat{G}_{1}||\hat{G}_{2}=\{\hat{X}_{1}\hat{\otimes} \hat{X}_{2}, \hat{\Sigma}_{1}\cup\hat{\Sigma}_{2}, \hat{\delta}_{1}||\hat{\delta}_{2}, \hat{x}_{01}\hat{\otimes} \hat{x}_{02}, \hat{x}_{1m}\hat{\otimes} \hat{x}_{2m} \}.
     \end{equation}
    Here $\hat{X}_{1}\hat{\otimes} \hat{X}_{2}=\{\hat{x}_{1}\hat{\otimes} \hat{x}_{2}, \hat{x}_{i}\in \hat{X}_{i}, i\in\{1, 2\} \}$, where $\hat{\otimes}$ denotes bi-fuzzy tensor operation. $\hat{\delta}_{1}||\hat{\delta}_{2}$ is a function from
    $(\hat{X}_{1}\hat{\otimes}\hat{X}_{2})\times(\hat{\Sigma}_{1}\cup\hat{\Sigma}_{2})$ to $(\hat{X}_{1}\hat{\otimes}\hat{X}_{2})$. That is, for any $\hat{x}_{1}\hat{\otimes} \hat{x}_{2}\in(\hat{X}_{1}\hat{\otimes}\hat{X}_{2})$ and any $\hat{\sigma}\in(\hat{\Sigma}_{1}\hat{\otimes}\hat{\Sigma}_{2})$,
    \begin{equation}
        (\hat{\delta}_{1}||\hat{\delta}_{2})(\hat{x}_{1}\hat{\otimes} \hat{x}_{2}, \hat{\sigma})=(\hat{x}_{1}\hat{\otimes} \hat{x}_{2})\hat{\odot}\hat{\sigma}.
    \end{equation}
    Here the corresponding matrix $\hat{\sigma}$ of bi-fuzzy event $\hat{\sigma}$ is defined as follows.
    \begin{enumerate}
    \item If bi-fuzzy event $\hat{\sigma}\in\hat{\Sigma}_{1}\cap\hat{\Sigma}_{2}$, then the matrix $\hat{\sigma} = \hat{\sigma}_{1}\hat{\otimes}\hat{\sigma}_{2}$, where $\hat{\sigma}_{1}$ and $\hat{\sigma}_{2} $ are the corresponding matrices of bi-fuzzy event $\hat{\sigma}$ in $\hat{G}_{1}$ and $\hat{G}_{2}$, respectively.
    \item If bi-fuzzy event $\hat{\sigma}\in\hat{\Sigma}_{1} \backslash \hat{\Sigma}_{2}$, then the matrix $\hat{\sigma} = \hat{\sigma}_{1}\hat{\otimes}\hat{I}_{2}$, where $\hat{\sigma}_{1}$ is the corresponding matrix of bi-fuzzy event $\hat{\sigma}$ in $\hat{G}_{1}$, and $\hat{I}_{2}$ is the unit matrix of order $|{X}_{2}|$.
    \item If bi-fuzzy event $\hat{\sigma}\in\hat{\Sigma}_{2} \backslash \hat{\Sigma}_{1}$, then the matrix $\hat{\sigma} = \hat{I}_{1}\hat{\otimes}\hat{\sigma}_{2}$, where $\hat{\sigma}_{2}$ is the corresponding matrix of bi-fuzzy event $\hat{\sigma}$ in $\hat{G}_{2}$, and $\hat{I}_{1}$ is the unit matrix of order $|{X}_{1}|$.
    \end{enumerate}
    As indicated above, the symbol $\hat{\otimes}$ denotes bi-fuzzy tensor of matrices. That is, for matrices $\hat{A}=[a_{ij}]_{i\in [1,m]}^{j\in [1,n]}$ and  $\hat{B}=[b_{pq}]_{p\in [1,k]}^{q \in [1,l]}$, $a_{ij}, b_{pq} \in NCFD$, we have\par
    \[
    \hat{A}\hat{\otimes} \hat{B} =
            \left[
                 \begin{array}{ccc}
                \tilde{a}_{11} \sqcap\hat{B} & {\ldots} & \tilde{a}_{1n} \sqcap \hat{B} \\
                \vdots & \ddots & \vdots\\
                \tilde{a}_{m1} \sqcap \hat{B} & {\ldots} & \tilde{a}_{mn} \sqcap \hat{B}
                \end{array}
            \right],\
    \]
    where
    \[
            \tilde{a}_{ij} \sqcap \hat{B} =
            \left[ \begin{array}{ccc}
                \tilde{a}_{ij}\sqcap \tilde{b}_{11} & \ldots & \tilde{a}_{ij}\sqcap \tilde{b}_{1l} \\
                \vdots & \ddots & \vdots\\
                \tilde{a}_{ij}\sqcap \tilde{b}_{k1} & \ldots & \tilde{a}_{ij}\sqcap \tilde{b}_{kl}
                \end{array}
            \right].
    \]
\par

   The following proposition states an important property concerning the bi-fuzzy language generated by the parallel composition of bi-fuzzy automata.

\begin{proposition}
  Given two bi-fuzzy automata $\hat{G}_{i}, i\in\{1, 2\}$, then
  \begin{equation}
   L(\hat{G}_{1}||\hat{G}_{2})(\hat{s}) = L(\hat{G}_{1})(\hat{s}) \sqcap L(\hat{G}_{2})(\hat{s}), \forall \hat{s} \in \hat{\Sigma}^{*}.
  \end{equation}

\end{proposition}

    \begin{IEEEproof}
      See Appendix B.
    \end{IEEEproof}

\section{Supervisory Control of Bi-Fuzzy DESs}

In this section, we focus on the supervisory control problem of BFDESs.  The controllability theorem and nonblocking controllability theorem are demonstrated. An algorithm  for checking the controllability conditions is presented.

\subsection{Controllability Theorem for Bi-Fuzzy DESs}
The uncontrolled  BFDES, usually called a plant, is modeled by a bi-fuzzy automaton $\hat{G}$. Suppose the behavior is not satisfactory and must be modified by a controller. Modifying the behavior could be understood as restricting the behavior to a subset of the bi-fuzzy language $L_{\hat{G}}$ by enabling the bi-fuzzy events with any fuzzy degree. However, each bi-fuzzy event $\hat{\sigma}\in\hat{\Sigma}$ is physically associated with a fuzzy degree of controllability. Then, we have the following notion.
\begin{definition}
  Uncontrollable events set $\hat{\Sigma}_{uc}$ and controllable events set $\hat{\Sigma}_{c}$ are defined as the functions
   from $ \hat{\Sigma} $ to $NCFD$,
    which satisfy the following condition:
  \begin{equation}
  \hat{\Sigma}_{c}(\hat{\sigma}) = \neg \hat{\Sigma}_{uc}(\hat{\sigma})  ~~(\forall \hat{\sigma} \in \hat{\Sigma}).
  \end{equation}
\end{definition}
\par
A supervisory controller, usually called as a bi-fuzzy supervisor, is a close-loop policy according to the observed system behavior dynamically.  Formally, the definition is given below.
\begin{definition}
A \emph{bi-fuzzy  supervisor} $\hat{S}$ is a close-loop control policy characterized by the following function:
\begin{equation}
\hat{S}:\hat{\Sigma}^{*}\rightarrow NCFD^{\hat{\Sigma}},
\end{equation}
where $NCFD^{\hat{\Sigma}}$ denotes the set of all functions from $\hat{\Sigma}$ to $NCFD$. For each $\hat{s}\in\hat{\Sigma}^{*}$ and each $\hat{\sigma} \in \hat{\Sigma}$, $\hat{S}(\hat{s})(\hat{\sigma})$ represents the fuzzy degree of the bi-fuzzy event $\hat{\sigma}$ being enabled after the occurrence of the bi-fuzzy events string $\hat{s}$.
\end{definition}
\par

In the light of the notion of admissible supervisors of DESs \cite{desbook} and FDES \cite{Qiu-1},  the fuzzy degree of $\hat{\sigma}$ following string $\hat{s}$  being physically  possible, together with the fuzzy degree of $\hat{\sigma}$ being uncontrollable, should be not greater than the fuzzy degree of $\hat{\sigma}$ being enabled by the supervisor $\hat{S}$ after $\hat{s}$ occurs. Thus, we introduce the following notion.
\begin{definition}
  A bi-fuzzy supervisor $\hat{S}$ is called as \emph{an admissible bi-fuzzy  supervisor } if the following condition holds.
  \begin{equation}
  \hat{\Sigma}_{uc}(\hat{\sigma})\sqcap  L_{\hat{G}}(\hat{s}\hat{\sigma}) \sqsubseteq \hat{S}(\hat{s})(\hat{\sigma}).
  \end{equation}
\end{definition}
Equation (20) is called as a \emph{bi-fuzzy admissibility condition} for bi-fuzzy supervisor $\hat{S}$ of $\hat{G}$.

\par
When an admissible bi-fuzzy supervisor $\hat{S}$ is ``attached" to an uncontrolled system $\hat{G}$, the controlled system will be generated,
which is characterized by the following notion.
\begin{definition}
  A \emph{bi-fuzzy controlled system} is denoted as $\hat{S}/\hat{G}$. The languages generated and marked by $\hat{S}/\hat{G}$, denoted by $L_{\hat{S}/\hat{G}}$ and $L_{\hat{S}/\hat{G}, m}$, respectively, are defined as follows: $L_{\hat{S}/\hat{G}}(\epsilon)= L_{\hat{S}/\hat{G},m}(\epsilon) = \frac{1}{1}$, and  $\forall \hat{s}\in\hat{\Sigma}^{*}$ with $|\hat{s}|>0$ and  $\forall \hat{\sigma} \in \hat{\Sigma}$,

      \begin{align}
        L_{\hat{S}/\hat{G}}(\hat{s}\hat{\sigma})&= L_{\hat{S}/\hat{G}}(\hat{s})\sqcap L_{\hat{G}}(\hat{s}\hat{\sigma})\sqcap  \hat{S}(\hat{s})(\hat{\sigma}),      \\
        L_{\hat{S}/\hat{G}, m}(\hat{s}) &= L_{\hat{S}/\hat{G}}(\hat{s}) \sqcap L_{\hat{S}/\hat{G}, m}(\hat{s}).
      \end{align}

\end{definition}
\par
A desired system behavior, usually called as a specification, in supervisory control is given by a bi-fuzzy language $\hat{K}$. $\hat{K}$ is called as a \emph{prefix-closed language} when $\hat{K}=pr(\hat{K})$. Here the ``$pr(\hat{K})$" is a function defined below.

\begin{definition}
  For any bi-fuzzy language $\hat{K}$ over $\hat{\Sigma}^{*}$, its \emph{prefix-closure} $pr(K):\hat{\Sigma}^{*}\rightarrow  NCFD $ is defined as:
  \begin{equation}
    pr(\hat{K})(\hat{s}) = \bigsqcup_{\hat{u} \in \hat{\Sigma}^{*}}\hat{K}(\hat{s} \cdot \hat{u}),
  \end{equation}
  where  $pr(\hat{K})(\hat{s})$ denotes the fuzzy degree of string $\hat{s}$ belonging to the prefix-closure of $\hat{K}$.
\end{definition}

\par
 Suppose that $L_{\hat{G}}$ is a bi-fuzzy language generated by a automaton, and $\hat{K}$, $\hat{K}_{1}$, $\hat{K}_{2}$ are bi-fuzzy languages over a common events set. Then, by means of  Proposition 1, Proposition 2, Proposition 3, and  Equation (23), the following properties concerning the $pr$ are easily obtained.
  \begin{equation}
  \  L_{\hat{G}} =  pr(L_{\hat{G}}).
  \end{equation}
  \begin{equation}
   \hat{K} \subseteq pr(\hat{K}).
  \end{equation}
  \begin{equation}
   \hat{K}_1 \subseteq \hat{K}_2 \Rightarrow pr(\hat{K}_1) \subseteq pr(\hat{K}_2).
  \end{equation}
\par
The objective of the supervisory control is to ensure that the controlled system is equivalent to the given specification. That is, $L_{\hat{S}/\hat{G}} = pr(\hat{K})$.
It should be pointed out that there might not exist a supervisor $\hat{S}$ that can  guarantee  $L_{\hat{S}/\hat{G}} = pr(\hat{K})$ for an arbitrary given specification $\hat{K}$.
However, what specifications are achievable? It is an interesting problem. The following theorem would discuss this problem.
\par

\begin{theorem}
  Let an uncontrolled BFDES be modeled by a bi-fuzzy automaton $\hat{G}=\{ \hat{X}, \hat{\Sigma}, \hat{\delta}$, $ \hat{x}_{0}\}$ with the bi-fuzzy uncontrollable events set $\hat{\Sigma}_{uc}\in NCFD^{\hat{\Sigma}}$. The specification is characterized by a bi-fuzzy language $\hat{K}$, which satisfies $\hat{K}\subseteq L_{\hat{G}}$ and $\hat{K}(\epsilon) = \frac{1}{1}$. Then there exists a bi-fuzzy supervisor $\hat{S}:\hat{\Sigma}^{*}\rightarrow NCVD^{\hat{\Sigma}}$ such that $\hat{S}$ satisfies the bi-fuzzy admissibility condition and $L_{\hat{S}/\hat{G}}=pr(\hat{K})$ if and only if for any $\hat{s} \in \hat{\Sigma}^{*}$ and any $\hat{\sigma} \in \hat{\Sigma}$,
  \begin{equation}
    pr(\hat{K})(\hat{s}) \sqcap \hat{\Sigma}_{uc}(\hat{\sigma}) \sqcap L_{\hat{G}}(\hat{s}\hat{\sigma}) \sqsubseteq pr(\hat{K})(\hat{s}\hat{\sigma}).
  \end{equation}
  Equation (27) is called as a \emph{bi-fuzzy controllability condition} of $\hat{K}$ with respect to $\hat{G}$ and $\hat{\Sigma}_{uc}$.
\end{theorem}

\begin{IEEEproof}
 We prove the sufficiency by constructing a bi-fuzzy supervisor as follows:
 \begin{equation}
    \hat{S}(\hat{s})(\hat{\sigma}) = pr(\hat{K})(\hat{s}\hat{\sigma}) \sqcup \hat{\Sigma}_{uc}(\hat{\sigma}),  \forall \hat{s}\in \hat{\Sigma}^{*}  \text{ and} \  \hat{\sigma} \in \hat{\Sigma}.
 \end{equation}
 \par
 It is easy to verify that the bi-fuzzy supervisor $\hat{S}$ satisfies the bi-fuzzy admissibility condition. We continue to show $L_{\hat{S}/\hat{G}} = pr(\hat{K})$  by induction on the length of bi-fuzzy events string $\hat{s}$.
 \par
 If $|\hat{s}| = 0$, i.e., $\hat{s}=\epsilon$, we have $L_{\hat{S}/\hat{G}}(\epsilon) = pr(\hat{K})(\epsilon) = \frac{1}{1}$. Suppose  $L_{\hat{S}/\hat{G}} = pr(\hat{K})$ holds when $|\hat{s}|  \leq n$. Then we need to show that $L_{\hat{S}/\hat{G}}(\hat{s}\hat{\sigma}) = pr(\hat{K})(\hat{s}\hat{\sigma}) , \forall \hat{\sigma} \in \hat{\Sigma}$ also holds.\par
 By the definition of $L_{\hat{S}/\hat{G}}$, we have
 \begin{align}
    L_{\hat{S}/\hat{G}}(\hat{s}\hat{\sigma}) = & L_{\hat{S}/\hat{G}}(\hat{s})\sqcap L_{\hat{G}}(\hat{s}\hat{\sigma})\sqcap  \hat{S}(\hat{s})(\hat{\sigma}) \nonumber \\
     = & L_{\hat{S}/\hat{G}}(\hat{s})\sqcap L_{\hat{G}}(\hat{s}\hat{\sigma}) \sqcap  (pr(\hat{K})(\hat{s}\hat{\sigma}) \sqcup
    \hat{\Sigma}_{uc}(\hat{\sigma})) \nonumber \\
     = & (L_{\hat{S}/\hat{G}}(\hat{s})\sqcap L_{\hat{G}}(\hat{s}\hat{\sigma}) \sqcap pr(\hat{K})(\hat{s}\hat{\sigma})) \sqcup \nonumber \\
       & (L_{\hat{S}/\hat{G}}(\hat{s}) \sqcap  L_{\hat{G}}(\hat{s}\hat{\sigma})   \sqcap    \hat{\Sigma}_{uc}(\hat{\sigma})). \nonumber
 \end{align}
 By means of  1) of Proposition 1, we have
  \[
  L_{\hat{S}/\hat{G}}(\hat{s})\sqcap L_{\hat{G}}(\hat{s}\hat{\sigma}) \sqcap pr(\hat{K})(\hat{s}\hat{\sigma}) \sqsubseteq  pr(\hat{K})(\hat{s}\hat{\sigma}).
  \]
  According to the given premiss $
  L_{\hat{S}/\hat{G}}(\hat{s})\sqcap L_{\hat{G}}(\hat{s}\hat{\sigma}) \sqcap    \hat{\Sigma}_{uc}(\hat{\sigma}) \sqsubseteq  pr(\hat{K})(\hat{s}\hat{\sigma})
  $, and  5) of Proposition 1, we obtain
  \begin{multline}
  L_{\hat{S}/\hat{G}}(\hat{s}\hat{\sigma})= (L_{\hat{S}/\hat{G}}(\hat{s})\sqcap L_{\hat{G}}(\hat{s}\hat{\sigma}) \sqcap   pr(\hat{K})(\hat{s}\hat{\sigma}))   \\
   \sqcup   (L_{\hat{S}/\hat{G}}(\hat{s})\sqcap L_{\hat{G}}(\hat{s}\hat{\sigma}) \sqcap    \hat{\Sigma}_{uc}(\hat{\sigma})) \sqsubseteq pr(\hat{K})(\hat{s}\hat{\sigma}).
  \end{multline}
  On the other hand, by virtue of  2) of Proposition 1, we have
  \begin{equation}
  pr(\hat{K})(\hat{s}\hat{\sigma})  \sqsubseteq pr(\hat{K})(\hat{s}\hat{\sigma}) \sqcup \hat{\Sigma}_{uc}(\hat{\sigma}) = \hat{S}(\hat{s})(\hat{\sigma}).
  \end{equation}
  According to the definition of $pr$, we have
  \begin{align*}
        pr(\hat{K})(\hat{s})  =& \big [ \bigsqcup_{\hat{u} \in \hat{\Sigma}^{*},\hat{\sigma}^{'} != \hat{\sigma}}(\hat{s}\hat{\sigma}^{'}\hat{u} )  \big ]  \sqcup   \big [ \bigsqcup_{\hat{u} \in \hat{\Sigma}^{*}}(\hat{s}\hat{\sigma}\hat{u} ) \big ]  \\
         =& \big [ \bigsqcup_{\hat{u} \in \hat{\Sigma}^{*},\hat{\sigma}^{'} != \hat{\sigma}}(\hat{s}\hat{\sigma}^{'}\hat{u} ) \big ] \sqcup \big [ pr(\hat{K})(\hat{s}\hat{\sigma}) \big ].
  \end{align*}
  Therefore, by  2) of Proposition 1, we have
  \begin{equation}
  pr(\hat{K})(\hat{s}\hat{\sigma}) \sqsubseteq pr(\hat{K})(\hat{s}).
  \end{equation}
  With $ \hat{K} \sqsubseteq  L_{\hat{G}}$, Equations (26) and (24), we have $ pr(\hat{K}) \sqsubseteq  pr(L_{\hat{G}}) = L_{\hat{G}}$. Thus, we have
  \begin{equation}
   pr(\hat{K})(\hat{s}\hat{\sigma}) \sqsubseteq  L_{\hat{G}}(\hat{s}\hat{\sigma}).
  \end{equation}
  Hence, from Equations (30) - (32) and  6) of Proposition 1, we have
  \begin{equation}
    pr(\hat{K})(\hat{s}\hat{\sigma}) \sqsubseteq (L_{\hat{G}}(\hat{s}\hat{\sigma}) \sqcap pr(\hat{K})(\hat{s}) \sqcap  \hat{S}(\hat{s})(\hat{\sigma})) =  L_{\hat{S}/\hat{G}}(\hat{s}\hat{\sigma}).
  \end{equation}
  With Equation (29) and Equation (33), we get
  \[
  pr(\hat{K})(\hat{s}\hat{\sigma}) = L_{\hat{S}/\hat{G}}(\hat{s}\hat{\sigma}).
  \]
  Therefore we have completed the proof of sufficiency. The remainder is the proof of necessity. That is, if $L_{\hat{S}/\hat{G}} = pr(\hat{K})$, we need to show
  $  pr(\hat{K})(\hat{s}) \sqcap \hat{\Sigma}_{uc}(\hat{\sigma}) \sqcap L_{\hat{G}}(\hat{s}\hat{\sigma}) \sqsubseteq pr(\hat{K})(\hat{s}\hat{\sigma})$.
  \par We have
  \begin{align}
           & pr(\hat{K})(\hat{s}) \sqcap \hat{\Sigma}_{uc}(\hat{\sigma}) \sqcap L_{\hat{G}}(\hat{s}\hat{\sigma}) \nonumber \\
        =  & pr(\hat{K})(\hat{s}) \sqcap L_{\hat{G}}(\hat{s}\hat{\sigma}) \sqcap (\hat{\Sigma}_{uc}(\hat{\sigma}) \sqcap L_{\hat{G}}(\hat{s}\hat{\sigma}))  \nonumber \\
        =  & L_{\hat{S}/\hat{G}}(\hat{s}) \sqcap L_{\hat{G}}(\hat{s}\hat{\sigma}) \sqcap (\hat{\Sigma}_{uc}(\hat{\sigma}) \sqcap L_{\hat{G}}(\hat{s}\hat{\sigma}))  \nonumber \\
        \sqsubseteq & L_{\hat{S}/\hat{G}}(\hat{s}) \sqcap L_{\hat{G}}(\hat{s}\hat{\sigma}) \sqcap \hat{S}(\hat{s})(\hat{\sigma}) \nonumber \\
        = & L_{\hat{S}/\hat{G}}(\hat{s}\hat{\sigma}).  \nonumber
  \end{align}
  Therefore, the theorem has been proved.
 \end{IEEEproof}

\par
Theorem 1 presents a necessary and sufficient condition for the existence of bi-fuzzy supervisors. Once the condition holds, the next important issue to be considered is the realization of the bi-fuzzy supervisor. It would be tedious or even impractical sometimes to list all $\hat{S}(\hat{s})$ for all $\hat{s}\in L_{\hat{G}}$ according to Equation (28). Hence, a more compact form of the supervisor is desired.
\par
Actually, the bi-fuzzy automaton that marks bi-fuzzy language $\hat{K}$ can serve as an automaton realization of the supervisor $\hat{S}$.
Let $\hat{R}$ be such an automaton:
$
\hat{R}=\{\hat{Q}, \hat{\Sigma}, \hat{\delta}, \hat{q}_{0}, \hat{q}_{m} \},
$
where $|Q|=n, \hat{q}_{m}=[\underbrace{\frac{1}{1}, \ldots, \frac{1}{1}}_{n}]$ and $ L_{m}(\hat{R}) = L(\hat{R}) = pr(\hat{K})$.
Then, by Proposition 4, we have the following equations.
\begin{multline}
L(\hat{G}||\hat{R}) = L(\hat{G}) {\cap} L(\hat{R})= L(\hat{G}){\cap}pr(\hat{K}) \\
=pr(\hat{K})=L(\hat{S}/\hat{G}).
\end{multline}
\begin{multline}
L_{m}(\hat{G}||\hat{R}) = L_{m}(\hat{G}){\cap} L_{m}(\hat{R})=L_{m}(\hat{G}){\cap}pr(\hat{K}) \\
=L_{m}(\hat{S}/\hat{G}).
\end{multline}
\par
That is, $\hat{G}||\hat{R}$ is exactly the behavior that is desired for the close-loop system $\hat{S}/\hat{G}$.
Therefore, in this sense, the supervisor $\hat{S}$ can be ``encoded" into a bi-fuzzy automaton and the control mechanism can be realized by the parallel composition of the uncontrolled system $\hat{G}$ and the supervisor automaton $\hat{R}$.

\subsection{An  Algorithm of Checking the Bi-Fuzzy Controllability Condition}
In this subsection, we present an algorithm to verify the bi-fuzzy controllability condition. An example is provided to illustrate the process of the method in detail. \par

For the sake of convenience, suppose $pr(\hat{K})$ is generated by a bi-fuzzy automaton $\hat{R}=\{\hat{Q}, \hat{\Sigma}, \hat{\delta}, \hat{q}_{0}, \hat{q}_{m}\}$, namely, $L_{\hat{R}} = pr(\hat{K})$. Then the bi-fuzzy controllability condition (Equation (27)) can be re-expressed as:
 \begin{equation}
 L_{\hat{R}}(\hat{s}) \sqcap \hat{\Sigma}_{uc}(\hat{\sigma}) \sqcap L_{\hat{G}}(\hat{s}\hat{\sigma}) \sqsubseteq L_{\hat{R}}(\hat{s}\hat{\sigma}), \ \forall   \hat{s} \in L_{\hat{R}}, \forall  \hat{\sigma} \in \hat{\Sigma}.
 \end{equation}
 \par
Intuitively, an exhaustive test could be made for each $\hat{s} $ and  each $\hat{\sigma}$ to verify the condition. However, the exhaustive test method is not feasible when the number of $\hat{s}$ is infinite. \par
Actually, it is not necessary to make an exhaustive test for each $\hat{s}$ because there might exist many different $\hat{s}_{i}$, such that all $L_{\hat{R}}(\hat{s}_i)$ are equal to each other, and  $L_{\hat{G}}(\hat{s}_i)$ as well. In this sense,  $[\hat{s}_i]$ could be called as an equivalent class. Then for an equivalent class $[\hat{s}_i]$, we only need to test the condition for only one $\hat{s}_i \in [\hat{s}_{i}]$.
Additionally, according to the definition of bi-fuzzy languages generated by automata (Equation (9)), we note that the values of $L_{\hat{G}}(\hat{s})$ and $L_{\hat{R}}(\hat{s})$ are only determined by the bi-fuzzy states transferring from the initial state driven by the occurrence of $\hat{s}$. Hence, an equivalent class can be represented by a pair of bi-fuzzy states in which the first and the second items are the bi-fuzzy states reachable from the inial states $\hat{x}_{0}$ and $\hat{q}_{0}$, respectively.
\par
Hence, the whole checking process can be divided into two steps as follows:
\begin{enumerate}
  \item Compute the set of all the accessible states pairs: $\{ ( \hat{x}_{0} \hat{\odot} \hat{s}, \hat{q}_{0} \hat{\odot} \hat{s}) | \hat{s} \in \hat{\Sigma}^{*}  \}$.

  \item Check the condition for each states pair and its following event $\hat{\sigma} \in \hat{\Sigma}$ successively until a violator is found. If a violator is found, then the controllability condition does not hold; otherwise, the controllability condition holds.
\end{enumerate}\par
 It should be pointed out that the accessible states pairs must be finite, otherwise, the two-step checking method is also unfeasible.
 Fortunately, from the definition of operation $\hat{\odot}$, we note that if $NCFD$  are specified with a finite $J$, then the accessible states of a bi-fuzzy automaton are finite, so is the accessible states pairs.\par

Qiu \cite{Qiu-1} presented a creative and effective method to get all the accessible states pairs based on the computing tree.
Actually, the basic idea of Qiu is inherited and utilized to solve our problem.
The computing tree is constructed as follows. The root is labelled by $(\hat{x}_0,\hat{q}_0)$. Each vertex, labelled by $(\hat{x}_0 \hat{\odot} \hat{s} ,\hat{q}_0\hat{\odot} \hat{s})$, may produce $n$'s sons vertices labelled by $(\hat{x}_0 \hat{\odot} \hat{s}\hat{\odot} \hat{\sigma}_{1} ,\hat{q}_0\hat{\odot} \hat{s}\hat{\odot} \hat{\sigma}_{1})$, $(\hat{x}_0 \hat{\odot} \hat{s}\hat{\odot} \hat{\sigma}_{2} ,\hat{q}_0\hat{\odot} \hat{s}\hat{\odot} \hat{\sigma}_{2})$, $\ldots$,
$(\hat{x}_0 \hat{\odot} \hat{s}\hat{\odot} \hat{\sigma}_{n} ,\hat{q}_0\hat{\odot} \hat{s}\hat{\odot} \hat{\sigma}_{n})$,  respectively. If a vertex whose label is equal to that of anther non-leaf vertex, then it is a leaf and marked by an underline. The computing ends with a leaf at the end of each branch.  Clearly, the labels of the tree vertices contain all the accessible states pairs \cite{Qiu-1}.\par

\begin{example}
  Let the plant  $\hat{G}=\{ \hat{X}, \hat{\Sigma}, \hat{\delta}, \hat{x}_{0}\}$ and  the specification $\hat{R}=\{\hat{Q}, \hat{\Sigma}, \hat{\delta}, \hat{q}_{0}\}$, where the common events $\hat{\Sigma} = \{ \hat{\sigma}_1,\hat{\sigma}_2\}$ with $\hat{\Sigma}_{uc}(\hat{\sigma}_{1})=\frac{1}{0.9}$, and $\hat{\Sigma}_{uc}(\hat{\sigma}_{2})=\frac{1}{0.1}$   and
     \[
   \hat{x}_{0}=[\frac{1}{1}~\frac{1}{0}],\  \hat{\sigma}_{1}^{1}=\left[
                                                \begin{array}{cc}
                                                   \frac{1}{0.6}+\frac{0.6}{0.9} & \frac{1}{0.9}+\frac{0.8}{1} \\
                                                   \frac{1}{0.3}+\frac{0.7}{0.6} & \frac{1}{0.3}+\frac{0.7}{0.6}  \\
                                                \end{array}
                                             \right],\
   \]
   \[
                                             \hat{\sigma}_{2}^{1}=\left[
                                                \begin{array}{cc}
                                                   \frac{1}{0.6}+\frac{0.6}{0.9} & \frac{1}{0.3}+\frac{0.7}{0.6} \\
                                                   \frac{1}{0.9}+\frac{0.8}{1}  & \frac{1}{0.6}+\frac{0.6}{0.9} \\
                                               \end{array}
                                             \right],
   \]\
   \[
   \hat{q}_{0}=[\frac{1}{1}~\frac{1}{0}], \ \hat{\sigma}_{1}^{2}=\left[
                                                \begin{array}{ccc}
                                                   \frac{1}{0.3}+\frac{0.7}{0.6} & \frac{1}{0.9}+\frac{0.8}{1} \\
                                                   \frac{1}{0.3}+\frac{0.7}{0.6} & \frac{1}{0.3}+\frac{0.7}{0.6}  \\
                                                \end{array}
                                             \right],\
   \]
   \[
                                             \hat{\sigma}_{2}^{2}=\left[
                                                \begin{array}{ccc}
                                                   \frac{1}{0.3}+\frac{0.7}{0.6} & \frac{1}{0.3}+\frac{0.7}{0.6} \\
                                                   \frac{1}{0.9}+\frac{0.8}{1}  & \frac{1}{0.6}+\frac{0.6}{0.9} \\
                                                \end{array}
                                             \right],
  \]
where $\hat{\sigma}_{1}^{1}, \hat{\sigma}_{2}^{1}$ and $\hat{\sigma}_{1}^{2}, \hat{\sigma}_{2}^{2}$ are the corresponding matrices of events $\hat{\sigma}_{1}$ and $\hat{\sigma}_{2}$  in $\hat{G}$ and $\hat{R}$, respectively. \par
Step 1: We obtain the computing tree first (as shown in Fig. 2), and get the all accessible bi-fuzzy states pairs (as shown in Table  \uppercase\expandafter{\romannumeral3}).\par
\begin{figure}
\center

\centering
\includegraphics[width=0.35\textwidth]{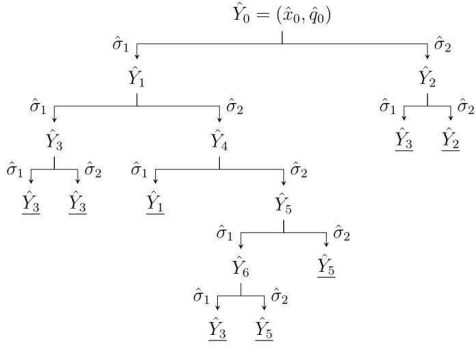}

\caption{\label{fig3} The computing tree is constructed for computing the accessible states of the bi-fuzzy automaton $\hat{G} \times \hat{R}$. The states $\hat{Y}_{i}\  ( i \in [0,6] )$ are listed in Table  \uppercase\expandafter{\romannumeral3}.}
\end{figure}

\begin{table*}[!htp]
    \center
    \caption{The bi-fuzzy states pairs $\hat{Y}_{i} ( i \in [0,6] )$ in Fig 3}
    \begin{tabular}{c|c|c}
        \hline
         $i$  &   $[\hat{s}]$   &   $\hat{Y}_{i}=(\hat{x}_{0} \hat{\odot} \hat{s} , \hat{q}_{0} \hat{\odot} \hat{s}) $ \\
        \hline
        $0$ & $[\epsilon]$ &  $([\frac{1}{1},\frac{1}{0}],[\frac{1}{1},\frac{1}{0}])$  \\
        \hline
        $1$ & $[\hat{\sigma}_{1}]$ &  $([\frac{1}{0.6}+\frac{0.6}{0.9},\frac{1}{0.9}+\frac{0.8}{1}],[\frac{1}{0.3}+\frac{0.7}{0.6},\frac{1}{0.9}+\frac{0.8}{1}])$  \\
        \hline
        $2$ & $[\hat{\sigma}_{2}]$ &  $([\frac{1}{0.6}+\frac{0.6}{0.9},\frac{1}{0.3}+\frac{0.7}{0.6}],[\frac{1}{0.3}+\frac{0.7}{0.6},\frac{1}{0.3}+\frac{0.7}{0.6}])$  \\
        \hline
        $3$ & $[\hat{\sigma}_{1}\hat{\sigma}_{1}]$ &  $([\frac{1}{0.6}+\frac{0.6}{0.9},\frac{1}{0.6}+\frac{0.6}{0.9}],[\frac{1}{0.3}+\frac{0.7}{0.6},\frac{1}{0.3}+\frac{0.7}{0.6}])$  \\
        \hline
        $4$ & $[\hat{\sigma}_{1}\hat{\sigma}_{2}]$ &
        $([\frac{1}{0.9}+\frac{0.8}{1},\frac{1}{0.6}+\frac{0.6}{0.9}],[\frac{1}{0.9}+\frac{0.8}{1},\frac{1}{0.6}+\frac{0.6}{0.9}])$  \\
        \hline
        $5$ & $[\hat{\sigma}_{1}\hat{\sigma}_{2}\hat{\sigma}_{2}]$ &
        $([\frac{1}{0.6}+\frac{0.6}{0.9},\frac{1}{0.6}+\frac{0.6}{0.9}],[\frac{1}{0.6}+\frac{0.6}{0.9},\frac{1}{0.6}+\frac{0.6}{0.9}])$  \\
        \hline
        $6$ & $[\hat{\sigma}_{1}\hat{\sigma}_{2}\hat{\sigma}_{2}\hat{\sigma}_{1}]$ &
        $([\frac{1}{0.6}+\frac{0.6}{0.9},\frac{1}{0.6}+\frac{0.6}{0.9}],[\frac{1}{0.3}+\frac{0.7}{0.6},\frac{1}{0.6}+\frac{0.6}{0.9}])$  \\
        \hline
    \end{tabular}
\end{table*}

Step 2: We test the condition (Equation (36)) for all the $\hat{s}$ in Table  \uppercase\expandafter{\romannumeral3} and all $\hat{\sigma} \in \hat{\Sigma}$. For the sake of convenience, we list all the cases in Table  \uppercase\expandafter{\romannumeral4}. \par

        \begin{table*}[!htp]
            \center
            \caption{Checking whether or not the Bi-Fuzzy Controllability Condition holds in Example 2}
            \begin{tabular}{c|c|c|c|c|c|c} \hline
            $\hat{s}$ & $\hat{\sigma}$ & $ L_{\hat{G}}(\hat{s}\hat{\sigma})=\hat{x}_{0}\hat{\odot} \hat{s} \hat{\odot}\hat{ \sigma }$ & $ L_{\hat{R}}(\hat{s})=\hat{q}_{0}\hat{\odot} \hat{s} $ &  $\hat{\Sigma}_{uc}(\hat{\sigma})$  &  $ L_{\hat{R}}(\hat{s}\hat{\sigma})=\hat{q}_{0}\hat{\odot} \hat{s} \hat{\odot}\hat{\sigma}$  &
            Equation (36) \\ \cline{1-7}

            \multirow{2}{*}{$\epsilon$} & $\hat{\sigma}_{1}$ & $\frac{1}{0.9}+\frac{0.8}{1}$ & $\frac{1}{1}$ & $\frac{1}{0.9}$ & $\frac{1}{0.9}+\frac{0.8}{1}$ & true  \\  \cline{2-7}

             & $\hat{\sigma}_{2}$ & $\frac{1}{0.6}+\frac{0.6}{0.9}$ & $\frac{1}{1}$ & $\frac{1}{0.1}$ & $\frac{1}{0.3}+\frac{0.7}{0.6}$ & true  \\  \cline{1-7}

            \multirow{2}{*}{$\hat{\sigma}_{1}$} & $\hat{\sigma}_{1}$ & $\frac{1}{0.6}+\frac{0.6}{0.9}$ & $\frac{1}{0.9}+\frac{0.8}{1}$ & $\frac{1}{0.9}$ & $\frac{1}{0.3}+\frac{0.7}{0.6}$ & false  \\  \cline{2-7}

             & $\hat{\sigma}_{2}$ & $\ldots$ & $\ldots$ & $\ldots$ & $\ldots$  & $\ldots$  \\  \cline{1-7}

            \end{tabular}
        \end{table*}

Table  \uppercase\expandafter{\romannumeral4} shows that the bi-fuzzy controllability condition dose not hold when  $\hat{s}=\hat{\sigma}_{1}$ and $\hat{\sigma} = \hat{\sigma}_{1}$. Hence, the language $L_{\hat{H}}$ is bi-fuzzy uncontrollable.\par

However, suppose that we specify another uncontrollable event function as follows:
$ \hat{\Sigma}_{uc}(\hat{\sigma}_{1}) = \frac{1}{0.1} \text{ and} \ \ \hat{\Sigma}_{uc}(\hat{\sigma}_{2}) = \frac{1}{0.9}$.
Following the same computing steps, we can obtain that the bi-fuzzy language $L_{\hat{H}}$ is bi-fuzzy controllable.
Then, as mentioned in subsection-(A), the $\hat{H}$ could serve as an automaton realization of the bi-fuzzy supervisor $\hat{S}$.
\end{example}

\subsection{Nonblocking Controllability Theorem for Bi-Fuzzy DESs}

In this subsection, we study the nonblocking supervisory control problem of BFDESs, which further requires the controlled BFDESs are nonblocking systems.  \par
Nonblocking is an important property of systems. The property requires a system should evolve without deadlock. Formally, the definition is given below.
\begin{definition}
  A BFDES $\tilde{G}=\{\tilde{X}, \tilde{\Sigma}, \tilde{\delta}, \tilde{x}_{0}, \tilde{x}_{m}\}$ is called as a \emph{nonblocking system} if and only if
  \begin{equation}
    L_{\hat{G}} = pr(L_{\hat{G},m}).
  \end{equation}
  A supervisor $\hat{S}$ is called as \emph{ nonblocking supervisor}, if and only if the controlled BFDES is a nonblocking system, i.e., $L_{\hat{S}/\hat{G}} = pr(L_{\hat{S}/\hat{G},m})$.
\end{definition}

 The following theorem would discuss  what specifications can be achieved by nonblocking supervisory control.
 \begin{theorem}
  Let an uncontrolled BFDES be modeled by a bi-fuzzy automaton $\hat{G}=\{ \hat{X}, \hat{\Sigma}, \hat{\delta}$, $\hat{x}_{0},  \hat{x}_{m}\}$ with the bi-fuzzy uncontrollable events set $\hat{\Sigma}_{uc}\in NCFD^{\hat{\Sigma}}$. The specification is characterized by a bi-fuzzy language $\hat{K}$, which satisfies $\hat{K}\subseteq L_{\hat{G,m}}$ and $\hat{K}(\epsilon) = \frac{1}{1}$. Then there exists a nonblocking bi-fuzzy supervisor $\hat{S}:\hat{\Sigma}^{*}\rightarrow NCVD^{\hat{\Sigma}}$ such that $\hat{S}$ satisfies the bi-fuzzy admissibility condition and
  \[
     L_{\hat{S}/\hat{G}} = pr(\hat{K}) \ \text{and}\ \  L_{\hat{S}/\hat{G},m} = \hat{K}
  \]
  if and only if $\hat{K}$ satisfies the bi-fuzzy controllability condition with respect to $\hat{G}$ and $\hat{\Sigma}_{uc}$, and
   $\hat{K}$ is $L_{m}(\hat{G})$-closure, that is, $\hat{K} = pr(\hat{K}) \cap L_{\hat{G},m}$.
\end{theorem}
\begin{IEEEproof}
For sufficiency, the proof of the $L_{\hat{S}/\hat{G}} = pr(\hat{K})$ is identical to that in Theorem 1. Hence, it is omitted here.
   Furthermore, we need to show $L_{\hat{S}/\hat{G},m} = \hat{K}$.
   Since $L_{\hat{S}/\hat{G}} = pr(\hat{K})$ has been shown, we have
  \[
       L_{\hat{S}/\hat{G},m} = L_{\hat{S}/\hat{G}} \cap L_{\hat{G},m} = pr(\hat{K}) \cap L_{\hat{G},m} = \hat{K}.
  \]
Therefore, the proof of sufficiency is completed.\par
For necessity, the proof of  the  bi-fuzzy controllability condition is also similar to that in Theorem 1. Hence, it is omitted here.
The remainder is to show $\hat{K} = pr(\hat{K}) \sqcap L_{\hat{G},m}$. Since there exists an nonblocking bi-fuzzy supervisor $\hat{S}$ such that
\[
     L_{\hat{S}/\hat{G}} = pr(\hat{K}),\ \text{and}\ \  L_{\hat{S}/\hat{G},m} = \hat{K},
\]
 by the definition of $L_{\hat{S}/\hat{G},m}$, we have
\[
\hat{K} = L_{\hat{S}/\hat{G},m} = L_{\hat{S}/\hat{G}} \cap   L_{\hat{G},m} = pr(\hat{K}) \cap   L_{\hat{G},m}.
\]
Therefore, the proof of Theorem 2 is completed.
\end{IEEEproof}
\par
 Besides the bi-fuzzy controllability condition, the achievable specifications in nonblocking supervisory control should satisfy $L_{m}(\hat{G})$-closure condition: $\hat{K} = pr(\hat{K}) \cap L_{\hat{G},m}$.
\par
  Suppose $pr(\hat{K})$ is generated by a bi-fuzzy automaton $\hat{R}=\{\hat{Q}, \hat{\Sigma}, \hat{\delta}, \hat{q}_{0}, \hat{q}_{m}\}$, namely, $L_{\hat{R}} = pr(\hat{K})$. Then the $L_{m}(\hat{G})$-closure condition can be re-expressed as: $\  \text{for any} \   \hat{s} \in \hat{\Sigma}^{*}$,
  $L_{\hat{R},m}(\hat{s}) = L_{\hat{R}}(\hat{s}) \sqcap L_{\hat{G},m}(\hat{s})$, which is only determined by the bi-fuzzy states pairs: $\{ ( \hat{x}_{0} \hat{\odot} \hat{s}, \hat{q}_{0} \hat{\odot} \hat{s}) | \hat{s} \in \hat{\Sigma}^{*}  \}$. Therefore, the two-step checking method mentioned in subsection B can also be used to verify the $L_{m}(\hat{G})$-closure condition.

\section{BFDESs vs. FDESs: An Illustrative Example in Traffic Control }
In the section, we present an illustrative example concerning traffic signal control.
A BFDES-based approach and an FDES-based approach will be used to solve the problem, respectively, and their performances will be compared with each other.
\par
Traffic signal control problem is very complicated. To be convenient for illustrating, the problem will be simplified as far as possible. We only consider an isolated intersection with two traffics directions and without turning traffic. Meanwhile, we use the following simple control approach.
\begin{enumerate}
 \item
 The parameters of the control, the basic green time $t_{bsc}$ and the maximum allowable green time $\tilde{t}_{\max}$, are set.
  \item
  The supervisor assigns the right-of-way to the green phase for the time ${t}_{bsc}$ .
  \item
  When the green time is expired, the supervisor will decide whether to
   extend the current green phase or switch to the next phase, according to the sensors data and the history of decisions. If the decision is ``switching", then goto Step 5), otherwise Goto Step 4).
  \item
  The current green phase is extended to a given time ${t}_{ext}$. Goto Step 3).
  \item
  The current green phase is terminated. Then the right-of-way is assigned to the new green phase for the time ${t}_{bsc}$. Goto Step 3).
\end{enumerate}
\par

The decision model can be characterized by the BFDES $\hat{G}=\{\hat{X},\hat{\Sigma},\hat{\delta}\}$ with $X=\{s,e\}$, where $s$ ($e$) denotes the decision of ``switching" (``extending", respectively).
 Events set $\hat{\Sigma}=\{ \hat{\sigma}_1, \hat{\sigma}_2 \}$, where $\hat{\sigma}_1$ ( $\hat{\sigma}_2$ ) is the abstract event that drives the supervisor to make the decision of ``switching"(``extending", respectively). Transition function $\hat{\delta}$ is characterized by  Fig. 3.
 \par
 \begin{figure}
\centering
\includegraphics[width=0.15\textwidth]{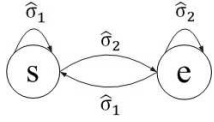}
\caption{\label{fig5}  The transition graph of decision model $\hat{G}$}
\end{figure}
\par

The uncontrollability function $\hat{\Sigma}_{uc}(\hat{\sigma})$, which could be thought of as the urgency of the corresponding decision, is usually given by a group of experts. In order to model the different opinions, the upper and lower boundaries of the functions should be provided. For instance, The uncontrollability function $\hat{\Sigma}_{uc}(\hat{\sigma})$ could be defined as follows.
\begin{subequations}
    \begin{gather}
        \begin{align}
            \hat{\Sigma}_{uc}(\hat{\sigma}_1) &=
                 \begin{cases}
                    \frac{1}{0},   & \text{if}\  t_{grn} \in [ 0, t_{bsc} ],\\
                   \min{ \{ \frac{({t}_{grn} - {t}_{bsc})^{2}}{({t}_{m} - {t}_{bsc})^{2}}, \frac{1}{1}}  \},
                                                                                                       & \text{if}\  {t}_{grn} \in ( t_{bsc} ,t_{maxu} ),\\
                    \frac{1}{1}  ,            & \text{if}\  {t}_{grn} \geq t_{maxu} , \\
                    \end{cases} \\
             \hat{\Sigma}_{uc}(\hat{\sigma}_2) &= \neg \hat{\Sigma}_{uc}(\hat{\sigma}_1).
        \end{align}
    \end{gather}
\end{subequations}
Here, ${t}_{m} \in [ t_{maxl} ,t_{maxu} ]$, and $[ t_{maxl} ,t_{maxu} ]$ = $Supp \{  \tilde{t}_{\max} \}$. $t_{grn}$ denotes the duration of the current green phase. We obtain the lower
(upper) boundary of the function when $t_m = t_{maxu}$ ($t_m = t_{maxl}$, respectively).

\par
 Both  traffics directions are equipped with a set of sensors, some sensors at the downstream for recording departure-vehicles, and the others at the upstream for arrival-vehicles.
  The supervisor will finally obtain the queue length of the current green and red phases, denoted by $\tilde{Q}_{grn}$ and $\tilde{Q}_{red}$, respectively.
  The data from the set of sensors will be synthesized to fuzzy quantities.
 \par
The queue length is an important parameter indicating traffic load. It is necessary to present an evaluating function to characterize the demand of the right-of-way based on the queue length. The evaluating function is usually given by a group of experts.
In order to model the different opinions, the upper and lower boundaries of the functions should be provided.
For instance, the following $\mu_{\hat{E}}(x)$ can serve as the evaluation function.

 \begin{equation}
 \mu_{\hat{E}}(x) =
 \begin{cases}
   \frac{1}{0}    ,                                                              &  \text{\ if \ }  x <= 0 , \\
    \exp (-\frac{(x-Q_{\max})^{2}}{2 * \sigma^{2}}) ,  & \text{\ if \ } x \in (0, Q_{\max}),\\
   \frac{1}{1}     ,                                                           &   \text{\ if \ }  x \geq Q_{\max},\\
 \end{cases}
 \end{equation}
 where $ \sigma \in [\sigma_l,\sigma_h]$, and $x$ is the queue length. $Q_{\max}$ is the end scale value of the sensors.
 $\mu_{\hat{E}}(x)$ with $\sigma = \sigma_l$ ($\sigma = \sigma_h$) is the lower (upper, respectively) boundary of the functions.
 Actually, the evaluation function $\mu_{\hat{E}}(x)$ could be viewed as the membership function of an interval type-2 fuzzy set $\hat{E}$. Then,
 the degrees of the demands of the right-of-way for the current green and red phases, denoted by $\tilde{D}_{grn}$ and $\tilde{D}_{red}$, respectively, are defined as follows.
 \begin{align}
   \tilde{D}_{grn} =& \hat{E} \cap \tilde{Q}_{grn},\\
   \tilde{D}_{red} =& \hat{E} \cap \tilde{Q}_{red},
 \end{align}
 where  $\cap$ is the fuzzy intersection operator \cite{Mendel-1}.\par
 Now, we can construct the matrices of the events as follows.
 \begin{subequations}
    \begin{gather}
        \begin{align}
           \hat{\sigma}_{1} &= \left[
                                     \begin{array}{cc}
                                        \tilde{D}_{grn} \sqcap \hat{\Sigma}_{uc}(\hat{\sigma}_{1})  &  \frac{1}{0} \\
                                        \tilde{D}_{grn} \sqcap \hat{\Sigma}_{uc}(\hat{\sigma}_{1})  &  \frac{1}{0} \\
                                    \end{array}
                              \right] ; \\
           \hat{\sigma}_{2} &= \left[
                                    \begin{array}{cc}
                                         \frac{1}{0}  \ \ &  \tilde{D}_{red} \sqcap \hat{\Sigma}_{uc}(\hat{\sigma}_{2}) \\
                                         \frac{1}{0}  \ \ &  \tilde{D}_{red} \sqcap \hat{\Sigma}_{uc}(\hat{\sigma}_{2}) \\
                                    \end{array}
                              \right].
        \end{align}
     \end{gather}
 \end{subequations}
\par
Noted that the events $\hat{\sigma}_{1}$ and $\hat{\sigma}_{2}$ occur simultaneously. Hence, suppose the current fuzzy state is $\hat{q}$, then the next fuzzy state $\hat{q}^{'} = [ \tilde{q}^{'}_{1} \ \  \tilde{q}^{'}_{2}]$ can be calculated as follows.
\begin{equation}
  \hat{q}^{'} = (\gamma_1 * (\hat{q} \hat{\odot} \hat{\sigma}_{1})) \sqcup (\gamma_2 * (\hat{q} \hat{\odot} \hat{\sigma}_{2})),
\end{equation}
where $\gamma_1$ and $\gamma_2$ are the weights of corresponding decisions.
Intuitively, the $\tilde{q}^{'}_{1}$  and $\tilde{q}^{'}_{2}$ could be regarded as the activation levels of the ``switching" decision and ``extending" decision, respectively.
Therefore, the supervisor will make a decision as follows.
\begin{equation}
 \begin{cases}
    \hat{S}_{s}(\hat{\sigma}_1) = \frac{1}{1} \text{ and } \hat{S}_{s}(\hat{\sigma}_2) = \frac{1}{0}       & \text{if}\ \ \ \ \ \   \tilde{q}^{'}_{1} \succeq  \tilde{q}^{'}_{2},\\
    \hat{S}_{s}(\hat{\sigma}_1) = \frac{1}{0} \text{ and } \hat{S}_{s}(\hat{\sigma}_2) = \frac{1}{1}       & \text{if}\ \ \ \ \ \   \tilde{q}^{'}_{1} \prec  \tilde{q}^{'}_{2}.\\
    \end{cases}
\end{equation}
Here the symbol $\succeq$ ($\prec$) denotes ``not less" (``less", respectively) relation in fuzzy theory, according to a certain fuzzy quantity ranking method.
\par

Following the same control process, similarly we can provide an FDES-based control approach,
which uses FDESs to characterize the decision model and the supervisor.
Due to the limitations of the FDES model, the $\tilde{t}_{\max}$ must be crisp numbers; and the uncontrollability function $\hat{\Sigma}_{uc}(\hat{\sigma})$ and the evaluating function $\mu_{\hat{E}}$ must be determinate functions; in addition, the sensors data $\tilde{Q}_{grn}$ and $\tilde{Q}_{red}$  must be synthesized to crisp numbers.
Intuitively, the FDES-based approach might lose many uncertainties due to the premature defuzzification to the original fuzzy data.
\par
At the end of this section, the two proposed approaches are implemented on simulation.
For simplicity, we use crisp intervals to denote the above mentioned fuzzy quantities.
\par

In the BFDES-based approach, the sensors data $\tilde{Q}_{grn}$ and $\tilde{Q}_{red}$ are simulated by adding a disturbance term to the exact data as follows.
\begin{align}
  \tilde{Q}_{grn} &=[Q_{grn}*(1-r_1*0.1) ,  Q_{grn}*(1+r_2*0.1)],\\
  \tilde{Q}_{red} &=[Q_{red}*(1-r_3*0.1) ,  Q_{red}*(1 + r_4*0.1)],
\end{align}
where $r_i, i\in[1,4]$ is a random value in [0,1], ${Q}_{red}$ and ${Q}_{grn}$ are the exact data.
\par
Similarly, in the FDES-based approach, the $\tilde{Q}_{grn}$ and $\tilde{Q}_{red}$ are simulated as follows.
\begin{align}
  \tilde{Q}_{grn} &= Q_{grn}*(1 + r_1 * 0.1),\\
  \tilde{Q}_{red} &= Q_{red}*(1 + r_2 * 0.1),
\end{align}
where $r_1$ and $r_2$ are random values in $[-1, 1]$.
The other parameters  are listed in Table  \uppercase\expandafter{\romannumeral5}.
\begin{table*}[!htp]
 \center
 \caption{Simulation Parameters}
\begin{tabular}{c|c|c}

\hline
Parameters & Values for the BFDES-based approach & Values for the FDES-based approach\\
 \cline{1-3}
Duration of simulation (hours) & 2   & 2  \\
\cline{1-3}
The number of lanes per approach  &  2  &   2 \\
\cline{1-3}
 $\tilde{t}_{\max}$ (seconds) & [60,80]  &  70\\
\cline{1-3}
 ${t}_{bsc}$  (seconds)  & 30  &  30\\
\cline{1-3}
${t}_{ext}$ (seconds) &   3 & 3\\
 \cline{1-3}

 The uncontrollability function $\hat{\Sigma}_{uc}(\hat{\sigma})$  &  Eq. (38) &  Eq. (38) but $t_m = (t_{maxl} + t_{maxu})/2$.\\
 \cline{1-3}

Evaluating function $\mu_{\hat{E}}(x)$ & Eq. (39) ($\sigma_{l}=10$ , $\sigma_{h}=30$, $Q_{\max}=90$) & Eq. (39) ($\sigma_{l}=\sigma_{h}=20$, $Q_{\max}=90$) \\
 \cline{1-3}

 The weights of decisions & $\gamma_1 = \gamma_2 = 1 $   & $\gamma_1 = \gamma_2 = 1 $  \\
 \cline{1-3}

Saturation flow rate (vehicles/hour) &  2880  &   2880 \\
\cline{1-3}
Lost time (seconds/cicle) &  4  &   4 \\
\cline{1-3}
\end{tabular}

\end{table*}

\par
The average delay time of the vehicles ($D_{avg}$) is an important control performance index in traffic signal control, which can be calculated as follows.
\begin{align}
  &D_{red}(i) = \Sigma_{j=1}^{n_i} L_{red}^{i}(j).\\
  &D_{grn}(i) = \Sigma_{j=1}^{n_i} L_{grn}^{i}(j). \\
  &D_{avg} = \frac{\Sigma_{i=1}^{m}(D_{red}(i) + D_{grn}(i))}{Q_{total}}.
\end{align}
Here the $Q_{total}$ denotes the total number of the arrival vehicles. The $D_{red}(i)$ ($D_{grn}(i)$) denotes the total delay time in the red phase (green phase, respectively) during the $i$th cycles (assume $m$ cycles totally). The $L_{red}^{i}(j)$ ($L_{grn}^{i}(j)$) denotes the number of the waiting vehicles in the red phase (green phase, respectively) at the $j$th second (assume there are $n_i$ seconds during the $i$th cycle).
\par
The arrival times of vehicles are assumed to be uniformly distributed. The average delay time $D_{avg}$  under various average arrival rates are shown in Table \uppercase\expandafter{\romannumeral6}.
\begin{table}[htp]
 \center
 \caption{The average delay time $D_{avg}$ }
\begin{tabular}{c|c|c}
\hline
 Arrival rate  & $D_{avg}$ (BFDES Approach) & $D_{avg}$ (FDES Approach)  \\
 \cline{1-3}

1: \ \ \  720 & 15.54 & 16.46 \\
\cline{1-3}

2: \ \   1800 & 19.04 & 21.16 \\
\cline{1-3}

3: \ \   2480 & 30.11 & 34.30  \\
\cline{1-3}
\end{tabular}
\end{table}

\par
The average queue length is another important performance index in traffic signal control. Fig. 4 shows the average queue lengths  under the three average arrival rates in Table  \uppercase\expandafter{\romannumeral6}.

\begin{figure}
\center

\centering
\includegraphics[width=0.3\textwidth]{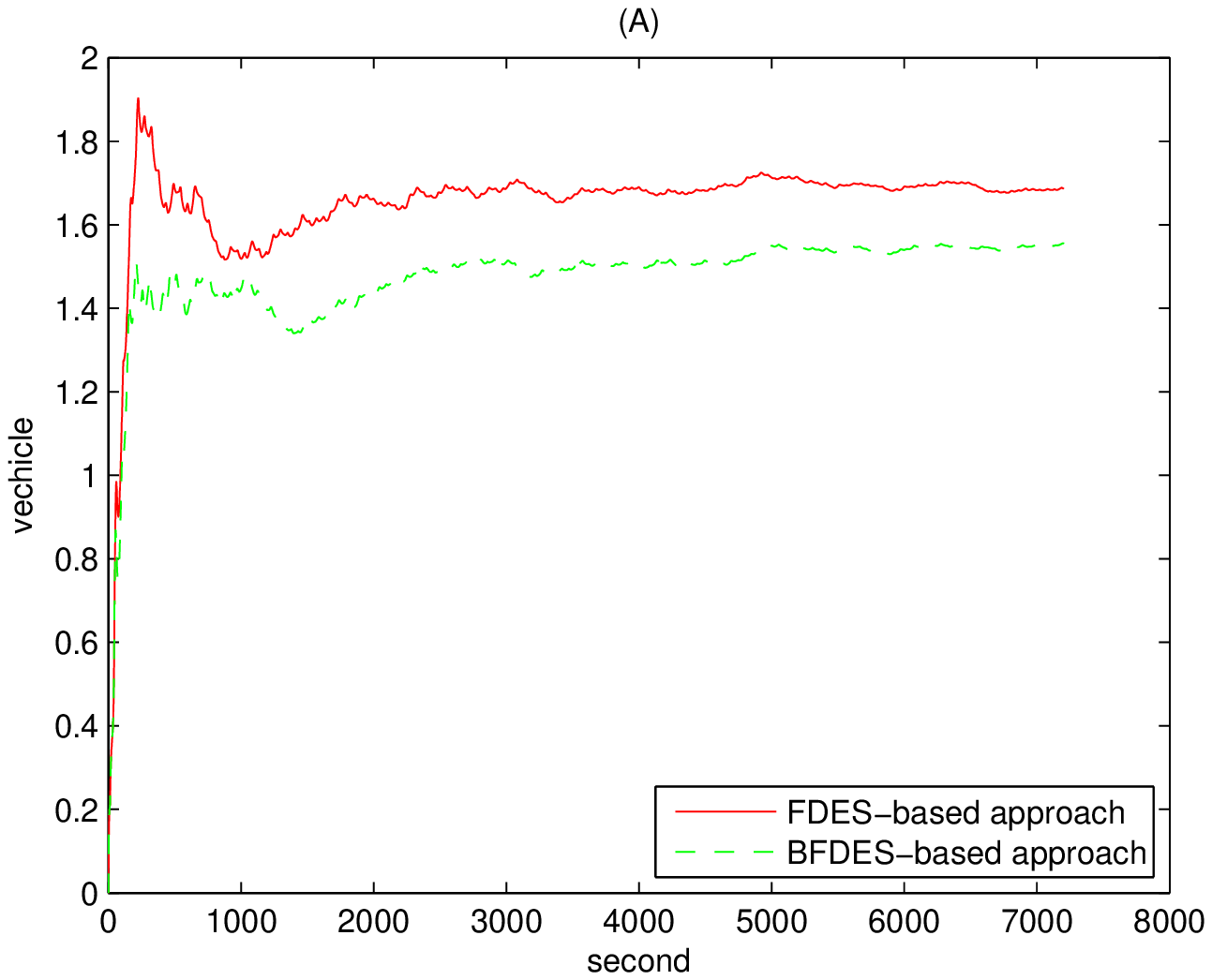}
\includegraphics[width=0.3\textwidth]{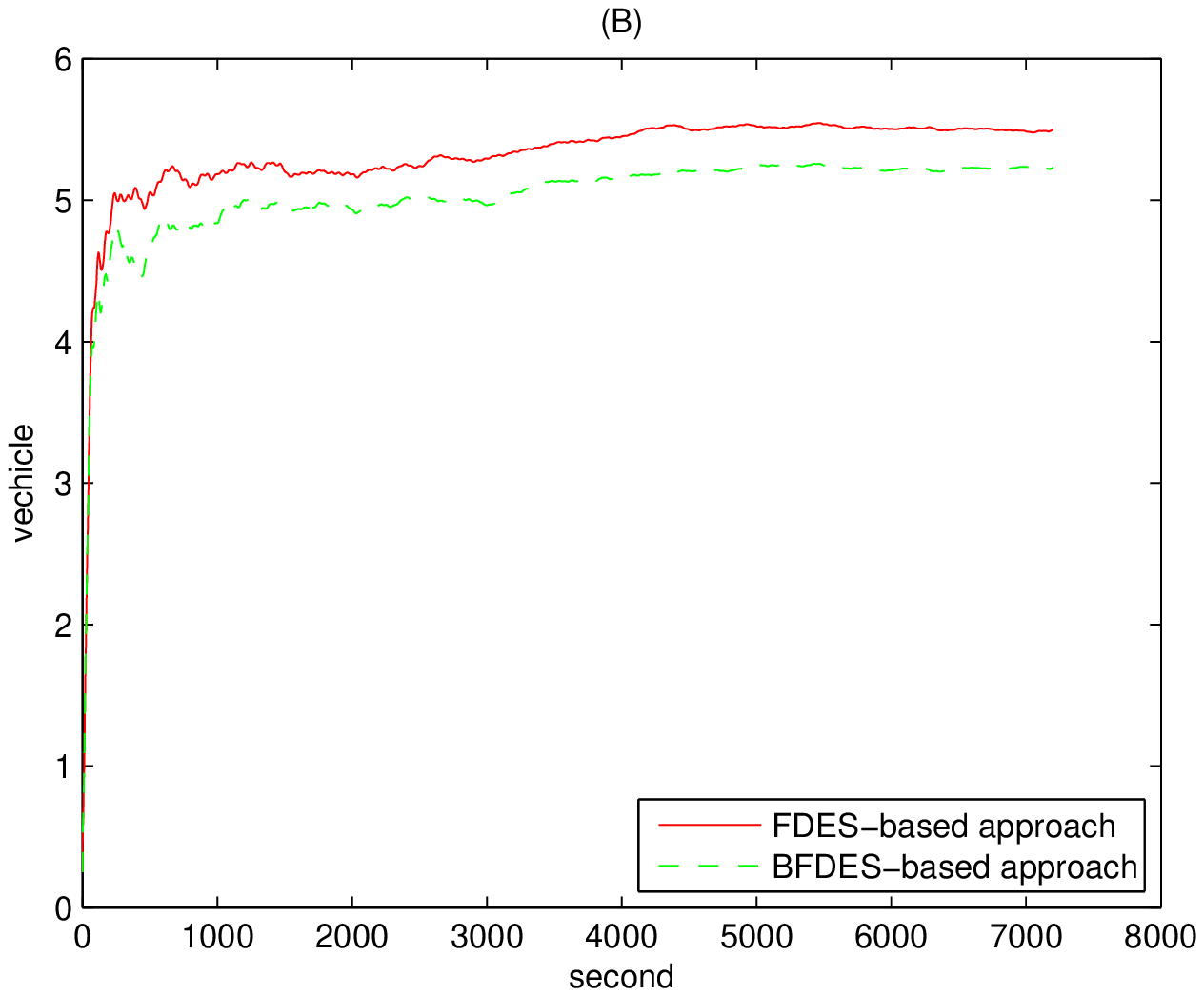}
\includegraphics[width=0.3\textwidth]{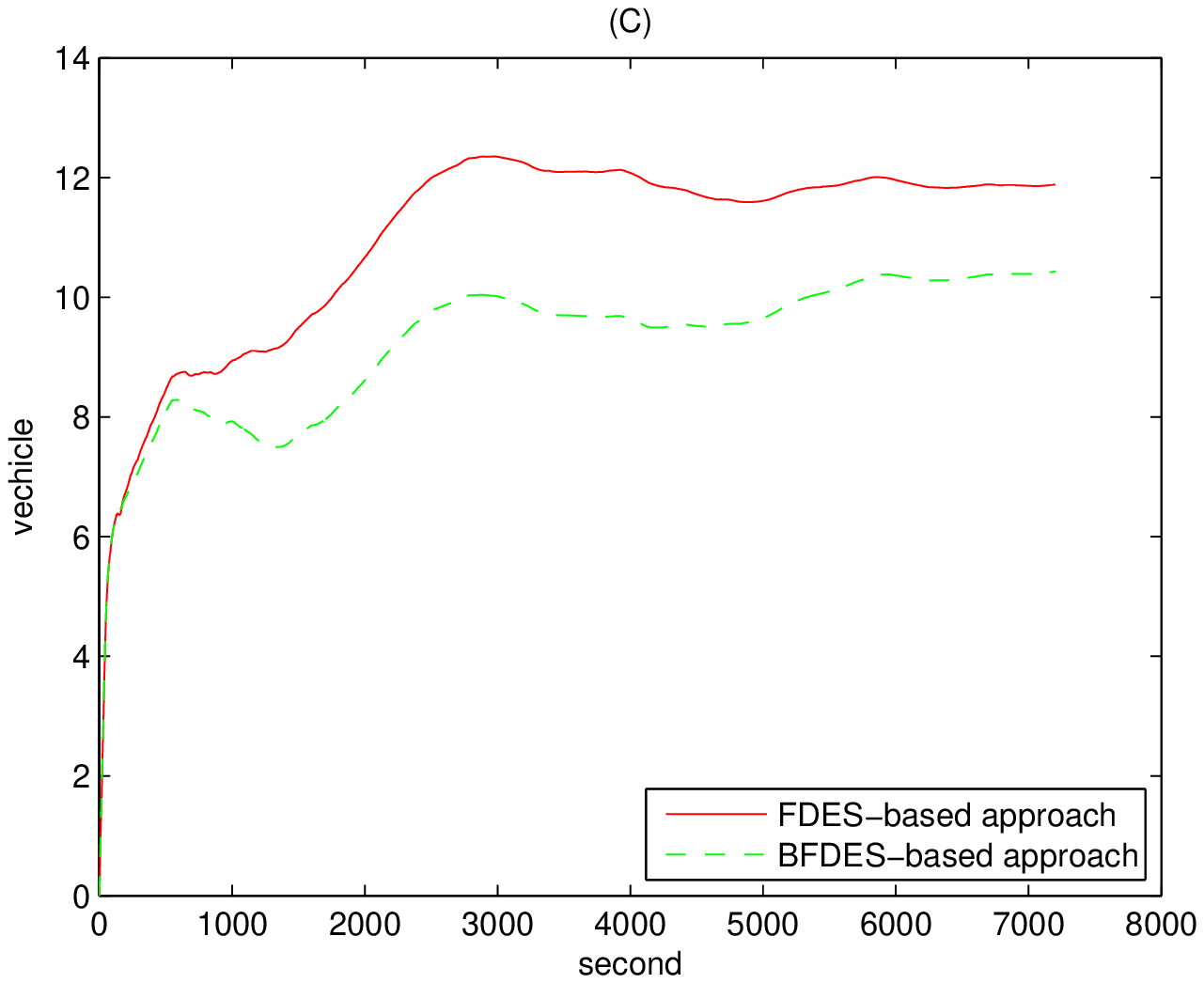}

\caption{
(A).the average queue length  under the first average arrival rate.
(B).the average queue length  under the second average arrival rate.
(C).the average queue length  under the third average arrival rate.
}
\end{figure}

Table \uppercase\expandafter{\romannumeral6} and Fig. 4. demonstrate that the BFDES-based approach has a better control performance than the FDES-based approach in general.
\par

\begin{remark}
BFDEs are constructed based upon T2 FSs.
Thus, BFDESs model can directly characterize some fuzzy and uncertain data of the physical systems.
Therefore, if BFDESs are used to model the physical systems, we do not have to defuzzify the original fuzzy data in the modeling phase.
In the above-mentioned example, we have directly modeled the different opinions of the experts by using BFDESs, but we need defuzzyify them if FDESs are used.
In addition, the experimental results reveal the fact that sometimes it is much better to process fuzzy data directly than to defuzzify them too early.
Hence, BFDESs  have well advantages over FDESs model in some cases.

\end{remark}

\section{Conclusions}
          FDESs were proposed by Lin and Ying \cite{Feng-1} based on T1 FSs theory. However, T1 FSs have limited capabilities to  handle directly  some  linguistic and data uncertainties. Thus, FDES [4-9] may
not be a so satisfactory model to characterize some high-uncertainty systems. To model higher-order uncertainties systems more precisely, a generalized FDESs model, namely BFDESs, have been formulated.
 The supervisory control theory of BFDESs has been developed. The controllability theorem and nonblocking controllability theorem have been demonstrated. Furthermore, an algorithm for checking the bi-fuzzy controllability condition has been introduced.
 The supremal controllable bi-fuzzy sublanguage and the infimal prefix-closed controllable bi-fuzzy superlanguage have been investigated in detail. The two controllable approximations to an uncontrollable language could be chosen to be the alternative schemes of the unachievable specifications in supervisory control.
  Finally, an example concerning traffic signal control has been provided to show the applicability and the advantages of BFDESs model.
  \par

\appendices
\section{controllable approximations to uncontrollable bi-fuzzy languages}
When a given specification cannot be achieved by supervisory control, it is naturally to consider getting an achievable approximation to the unachievable specification.
In this section, we will first investigate several basic properties concerning the controllability of bi-fuzzy languages, and then  consider how to get the best controllable approximations to an uncontrollable bi-fuzzy language.

 \par
 According to Theorem 1, the following notion is obtained directly.
\begin{definition}
  Let $\hat{K}$ and $\hat{M}$ be bi-fuzzy languages over bi-fuzzy events set $\hat{\Sigma}$ with $pr(\hat{M})=\hat{M}$ and $\hat{K} \subseteq \hat{M}$. $\hat{K}$ is said to be \emph{bi-fuzzy controllable} with respect to $\hat{M}$ and $\hat{\Sigma}_{uc}$ if  $\forall \hat{s}\in\hat{\Sigma}^{*}$ and $\forall \hat{\sigma}\in\hat{\Sigma}$,
  \begin{equation}
   pr(\hat{K})(\hat{s})\sqcap \hat{\Sigma}_{uc}(\hat{\sigma})\sqcap \hat{M}(\hat{s}\hat{\sigma})  \sqsubseteq  pr(\hat{K})(\hat{s}\hat{\sigma}).
  \end{equation}
\end{definition}
\par

 Clearly, if $\hat{K}$ is bi-fuzzy controllable, then so is $pr(\hat{K})$. The following proposition characterizes some fundamental properties of controllable bi-fuzzy languages.
\begin{proposition}
Let $\hat{K}_{1}$ and $\hat{K}_{2}$ be bi-fuzzy languages over bi-fuzzy events set $\hat{\Sigma}$. Then
  \begin{enumerate}
    \item if $\hat{K}_{1}$ and $\hat{K}_2$ are bi-fuzzy controllable, then $\hat{K}_1 \cup \hat{K}_2$ is bi-fuzzy controllable;
    \item if $pr(\hat{K}_1) = \hat{K}_1$ and $pr(\hat{K}_2) = \hat{K}_2$, then $pr(\hat{K}_1 \cup \hat{K}_2) = \hat{K}_1 \cup \hat{K}_2$;
    \item if $\hat{K}_{1}$ and $\hat{K}_2$ are bi-fuzzy controllable and $pr(\hat{K}_1) \cap pr(\hat{K}_2) = pr(\hat{K}_1 \cap \hat{K}_2)$, then $\hat{K}_1 \cap \hat{K}_2$ is bi-fuzzy controllable;
    \item if $pr(\hat{K}_i) = \hat{K}_i$, and $K_i$ is controllable, $i =\{1,2\}$, then $pr(\hat{K}_1) \cap pr(\hat{K}_2) = pr(\hat{K}_1 \cap \hat{K}_2)$ and $\hat{K}_1 \cap \hat{K}_2$ is bi-fuzzy controllable.
  \end{enumerate}
\end{proposition}
\begin{IEEEproof}
  1) For any $\hat{s} \in \hat{\Sigma}^{*}$ and any $\hat{\sigma} \in \hat{\Sigma}$, we have
  \begin{align}
        &pr(\hat{K}_1 \cup \hat{K}_2)(\hat{s}) \sqcap \hat{\Sigma}_{uc}(\hat{\sigma}) \sqcap \hat{M}(\hat{s}\hat{\sigma})  \nonumber \\
      = &\bigcup_{\hat{u} \in \hat{\Sigma}^{*}}(\hat{K}_1 \cup \hat{K}_2)(\hat{s}\cdot\hat{u}) \sqcap \hat{\Sigma}_{uc}(\hat{\sigma}) \sqcap \hat{M}(\hat{s}\hat{\sigma})   \nonumber \\
      = & \bigcup_{\hat{u} \in \hat{\Sigma}^{*}}(\hat{K}_1(\hat{s}\cdot\hat{u}) \sqcup \hat{K}_2(\hat{s}\cdot\hat{u})) \sqcap \hat{\Sigma}_{uc}(\hat{\sigma}) \sqcap \hat{M}(\hat{s}\hat{\sigma})    \nonumber \\
       = & \bigg[ \big [ \bigcup_{\hat{u} \in \hat{\Sigma}^{*}}\hat{K}_1(\hat{s}\cdot\hat{u}) \big ] \sqcup \big [  \bigcup_{\hat{u} \in \hat{\Sigma}^{*}} \hat{K}_2(\hat{s}\cdot\hat{u}) \big ] \bigg ] \sqcap \hat{\Sigma}_{uc}(\hat{\sigma}) \sqcap \hat{M}(\hat{s}\hat{\sigma})    \nonumber \\
      = & \big [ \bigcup_{\hat{u} \in \hat{\Sigma}^{*}} \hat{K}_1(\hat{s}\cdot\hat{u}) \sqcap \hat{\Sigma}_{uc}(\hat{\sigma}) \sqcap \hat{M}(\hat{s}\hat{\sigma}) \big ] \sqcup   \nonumber  \\
        & \big [ \bigcup_{\hat{u} \in \hat{\Sigma}^{*}}\hat{K}_2(\hat{s}\cdot\hat{u}) \sqcap \hat{\Sigma}_{uc}(\hat{\sigma}) \sqcap \hat{M}(\hat{s}\hat{\sigma}) \big ]    \nonumber \\
      = & \big [ pr(\hat{K}_1)(\hat{s}) \sqcap \hat{\Sigma}_{uc}(\hat{\sigma}) \sqcap \hat{M}(\hat{s}\hat{\sigma}) \big ] \sqcup \nonumber \\
        & \big [ pr(\hat{K}_2)(\hat{s}) \sqcap \hat{\Sigma}_{uc}(\hat{\sigma}) \sqcap \hat{M}(\hat{s}\hat{\sigma}) \big ]   \nonumber \\
      \sqsubseteq & pr(\hat{K}_1)(\hat{s}\hat{\sigma}) \sqcup pr(\hat{K}_2)(\hat{s}\hat{\sigma})  \ \nonumber \\
      = & \big [ \bigcup_{\hat{u} \in \hat{\Sigma}^{*}} \hat{K}_1 (\hat{s}\hat{\sigma} \cdot u ) \big ] \sqcup \big [ \bigcup_{\hat{u} \in \hat{\Sigma}^{*}} \hat{K}_2 (\hat{s}\hat{\sigma} \cdot u ) \big ]   \nonumber \\
       = &  \bigcup_{\hat{u} \in \hat{\Sigma}^{*}} \big [ \hat{K}_1 (\hat{s}\hat{\sigma} \cdot u )  \sqcup \hat{K}_2 (\hat{s}\hat{\sigma} \cdot u ) \big ]  \  \nonumber \\
      = & \bigcup_{\hat{u} \in \hat{\Sigma}^{*}} (\hat{K}_1 \cup \hat{K}_2 )(\hat{s}\hat{\sigma} \cdot u )    \nonumber \\
      = & pr(\hat{K}_1 \cup \hat{K}_2)(\hat{s} \cdot \hat{\sigma})     \nonumber
  \end{align}
  2) For any $\hat{s} \in \hat{\Sigma}^{*}$, we have
  \begin{align}
  pr(\hat{K}_1 \cup \hat{K}_2)(s)
      & = \bigcup_{\hat{u} \in \hat{\Sigma}^{*}}(\hat{K}_1 \cup \hat{K}_2)(\hat{s}\hat{u})      \nonumber \\
      & = \bigcup_{\hat{u} \in \hat{\Sigma}^{*}}\hat{K}_1(\hat{s}\hat{u}) \cup  \bigcup_{\hat{u} \in \hat{\Sigma}^{*}}\hat{K}_2(\hat{s}\hat{u})
             \nonumber \\
       & = pr(\hat{K}_1)(\hat{s}) \sqcup pr(\hat{K}_2)(\hat{s})     \nonumber \\
       & = \hat{K}_1(\hat{s}) \sqcup \hat{K}_2(\hat{s})     \nonumber \\
       & = (\hat{K}_1 \cup \hat{K}_2)(\hat{s})     \nonumber
  \end{align}
  3) For any $\hat{s} \in \hat{\Sigma}^{*}$ and any $\hat{\sigma} \in \hat{\Sigma}$, we have
  \begin{align}
        &pr(\hat{K}_1 \cap \hat{K}_2)(\hat{s}) \sqcap \hat{\Sigma}_{uc}(\hat{\sigma}) \sqcap \hat{M}(\hat{s}\hat{\sigma})  \nonumber \\
      = &\bigcup_{\hat{u} \in \hat{\Sigma}^{*}}(\hat{K}_1 \cap \hat{K}_2)(\hat{s}\cdot\hat{u}) \sqcap \hat{\Sigma}_{uc}(\hat{\sigma}) \sqcap \hat{M}(\hat{s}\hat{\sigma})   \nonumber \\
      = & \bigcup_{\hat{u} \in \hat{\Sigma}^{*}}(\hat{K}_1(\hat{s}\cdot\hat{u}) \sqcap \hat{K}_2(\hat{s}\cdot\hat{u})) \sqcap \hat{\Sigma}_{uc}(\hat{\sigma}) \sqcap \hat{M}(\hat{s}\hat{\sigma})     \nonumber \\
      \sqsubseteq  & \bigg[ \big [ \bigcup_{\hat{u} \in \hat{\Sigma}^{*}}\hat{K}_1(\hat{s}\cdot\hat{u}) \big ] \sqcap \big [  \bigcup_{\hat{u} \in \hat{\Sigma}^{*}} \hat{K}_2(\hat{s}\cdot\hat{u}) \big ] \bigg ] \sqcap \hat{\Sigma}_{uc}(\hat{\sigma}) \sqcap \hat{M}(\hat{s}\hat{\sigma})    \nonumber \\ 
      = & \big [ \bigcup_{\hat{u} \in \hat{\Sigma}^{*}} \hat{K}_1(\hat{s}\cdot\hat{u}) \sqcap \hat{\Sigma}_{uc}(\hat{\sigma}) \sqcap \hat{M}(\hat{s}\hat{\sigma}) \big ] \sqcap \nonumber \\
        & \big [ \bigcup_{\hat{u} \in \hat{\Sigma}^{*}}\hat{K}_2(\hat{s}\cdot\hat{u}) \sqcap \hat{\Sigma}_{uc}(\hat{\sigma}) \sqcap \hat{M}(\hat{s}\hat{\sigma}) \big ]    \nonumber \\
      = & \big [ pr(\hat{K}_1)(\hat{s}) \sqcap \hat{\Sigma}_{uc}(\hat{\sigma}) \sqcap \hat{M}(\hat{s}\hat{\sigma}) \big ] \sqcap \nonumber \\
        & \big [ pr(\hat{K}_2)(\hat{s}) \sqcap \hat{\Sigma}_{uc}(\hat{\sigma}) \sqcap \hat{M}(\hat{s}\hat{\sigma}) \big ]   \nonumber \\
      \sqsubseteq & pr(\hat{K}_1)(\hat{s}\hat{\sigma}) \sqcap pr(\hat{K}_2)(\hat{s}\hat{\sigma})   \nonumber \\
      = & (pr(\hat{K}_1) \cap pr(\hat{K}_2))(\hat{s}\hat{\sigma})    \nonumber \\
      = &  pr(\hat{K}_1 \cap \hat{K}_2)(\hat{s}\hat{\sigma}).      \nonumber
  \end{align}
  4). For any $\hat{s} \in \hat{\Sigma}^{*}$ and any $\hat{\sigma} \in \hat{\Sigma}$, we have
  \begin{flalign}
   pr(\hat{K}_1 \cap \hat{K}_2)(\hat{s})
    = & \bigcup_{\hat{u} \in \hat{\Sigma}^{*}} (\hat{K}_1 \cap \hat{K}_2) (\hat{s}\hat{u}) \  \nonumber \\
    = & \bigcup_{\hat{u} \in \hat{\Sigma}^{*}} (\hat{K}_1  (\hat{s}\hat{u}) \sqcap \hat{K}_2 (\hat{s}\hat{u})) \  \nonumber \\
    \sqsubseteq & \bigcup_{\hat{u} \in \hat{\Sigma}^{*}} (\hat{K}_1  (\hat{s}\hat{u})) \sqcap \bigcup_{\hat{u} \in \hat{\Sigma}^{*}} (\hat{K}_2 (\hat{s}\hat{u})) \ \nonumber \\
    = & pr(\hat{K}_1)  (\hat{s}) \sqcap pr(\hat{K}_2) (\hat{s}) \ \   \nonumber \\
    = & ( pr(\hat{K}_1) \cap pr(\hat{K}_2)) (\hat{s}) \ \    \nonumber \\
    = & (\hat{K}_1 \cap \hat{K}_2) (\hat{s}) \  \nonumber \\
    \sqsubseteq & pr(\hat{K}_1 \cap \hat{K}_2)(\hat{s}).  \nonumber
  \end{flalign}
  Therefore, we get $pr(\hat{K}_1) \cap pr(\hat{K}_2) = pr(\hat{K}_1 \cap \hat{K}_2)$. The remainder of the proof is identical to  3) of Proposition 5. Hence, it is omitted here.
\end{IEEEproof}\par
   Proposition 5 shows that the controllable languages are closed under the union operation, and the prefix-closed and controllable languages are closed under the intersection operation.
\par
In  practical applications, the specifications given by bi-fuzzy languages sometimes cannot be achieved by supervisory control, namely, the bi-fuzzy languages are uncontrollable. Under this circumstance, it is natural to consider getting a controllable approximation to an uncontrollable language. Based on the results from Proposition 5, the controllable languages could be divided into the following two sets.
\begin{definition}
  The set of \emph{controllable sub-languages} of $\hat{K}$ and the set of \emph{prefix-closed and controllable super-languages} of $\hat{K}$ are defined as follows, respectively.
  \begin{align}
    C_{sub}(\hat{K}) &= \{ \hat{L}  : \hat{L} \subseteq \hat{K}   \}, \\
    C_{sup}(\hat{K}) &= \{ \hat{L}  : \hat{K} \subseteq \hat{L} \subseteq \hat{M} \text{, and} \ \ \hat{L}=pr(\hat{L}) \}.
  \end{align}
  Here $\hat{L}$ is bi-fuzzy controllable Language.
\end{definition}
\par
In general, we are more interested in the ``biggest"  controllable sub-language and the ``smallest" controllable super-language of $\hat{K}$.
Then, we try to consider the following two languages derived from $C_{sub}(\hat{K})$ and $C_{sup}(\hat{K})$.
 \begin{definition}
  The \emph{supremal controllable sub-language} and the \emph{infimal prefix-closed and controllable super-language} of $\hat{K}$ are defined as follows, respectively.
  \begin{equation}
    \hat{K}^{\uparrow} = \bigcup_{\hat{L} \in C_{sub}(\hat{K})} \hat{L} ,
  \end{equation}
  and
  \begin{equation}
    \hat{K}^{\downarrow} = \bigcap_{\hat{L} \in C_{sup}(\hat{K})} \hat{L}.
  \end{equation}
\end{definition}
\par
It is easy to deduce that if $\hat{K}$ is bi-fuzzy controllable, then $\hat{K}^{\uparrow} = \hat{K} $. In addition, if $\hat{K}$ is prefix-close and bi-fuzzy controllable, then $\hat{K}^{\downarrow} = \hat{K} $.
\par
The following proposition shows that $\hat{K}^{\uparrow}$ is the ``biggest"  controllable sub-language and  $\hat{K}^{\downarrow}$ is the ``smallest" controllable super-language of $\hat{K}$ indeed.

\begin{proposition}
Suppose $\hat{K}$ is a bi-fuzzy language. Then \\
1) $  \forall \hat{L} \in C_{sub}(\hat{K}), \ \hat{L} \subseteq \hat{K}^{\uparrow}\ \text{and} \ \hat{K}^{\uparrow} \in C_{sub}(\hat{K});$ \nonumber \\
2) $ \forall \hat{L} \in C_{sup}(\hat{K}), \  \hat{K}^{\downarrow} \subseteq \hat{L}\ \text{and} \ \hat{K}^{\downarrow} \in C_{sup}(\hat{K})$. \nonumber
\end{proposition}
\begin{IEEEproof}
  1) For any $\hat{s} \in \hat{\Sigma}^{*} $, and $\hat{L} \in C_{sub}(\hat{K})$, by virtue of  2) of Proposition 1, we have $\hat{L}(\hat{s}) \sqsubseteq \bigcup_{\hat{L}^{'} \in C_{sub}(\hat{K})} \hat{L}^{'}(\hat{s}) = \hat{K}^{\uparrow}(\hat{s}) $. That is, $\hat{L} \subseteq \hat{K}^{\uparrow}$. The remainder is to show $\hat{K}^{\uparrow} \in C_{sub}(\hat{K})$. According to  1) of Proposition 5, we get $\hat{K}^{\uparrow}$ is bi-fuzzy controllable. Furthermore, from 5) of Proposition 1, we have $ \hat{K}^{\uparrow}(\hat{s}) = \bigcup_{\hat{L} \in C_{sub}(\hat{K})} \hat{L}(\hat{s}) \sqsubseteq \hat{K}(\hat{s}) $. That is, $\hat{K}^{\uparrow} \subseteq \hat{K} $. Hence, we obtain $\hat{K}^{\uparrow} \in C_{sub}(\hat{K}) .$
  \par
  2) For any $\hat{s} \in \hat{\Sigma}^{*} $, and $\hat{L} \in C_{sup}(\hat{K})$, by virtue of 1) of Proposition 1, we have $ \hat{K}^{\downarrow}(\hat{s}) = \bigcap_{\hat{L}^{'} \in C_{sup}(\hat{K})} \hat{L}^{'}(\hat{s}) \sqsubseteq \hat{L}(\hat{s}) $. That is, $\hat{K}^{\downarrow} \subseteq \hat{L} $. The remainder is to show $\hat{K}^{\downarrow} \in C_{sup}(\hat{K})$. According to  4) of Proposition 5, we get $\hat{K}^{\downarrow}$ is bi-fuzzy controllable and $\hat{K}^{\downarrow} = pr(\hat{K}^{\downarrow})$. Moreover, from 6) of Proposition 1, we have $\hat{K}(\hat{s}) \sqsubseteq \bigcap_{\hat{L} \in C_{sub}(\hat{K})} \hat{L}(\hat{s}) = \hat{K}^{\downarrow}(\hat{s}) \sqsubseteq \hat{M}(\hat{s})$. That is, $\hat{K} \subseteq \hat{K}^{\downarrow}\subseteq \hat{M}$. Hence, we obtain $\hat{K}^{\downarrow} \in C_{sup}(\hat{K}).$
\end{IEEEproof}
\par
Proposition 6 suggests that $\hat{K}^{\uparrow}$  and  $\hat{K}^{\downarrow}$ could be thought of as the best controllable approximations to an uncontrollable language  $\hat{K}$.
Thus, $\hat{K}^{\uparrow}$  and  $\hat{K}^{\downarrow}$ could be chosen as the alternative schemes if the given specification  $\hat{K}$ cannot be achieved by supervisory control.
\par

The following two propositions characterize some basic properties concerning $\hat{K}^{\uparrow}$ and $\hat{K}^{\downarrow}$.
\begin{proposition}
  Suppose $\hat{K}$, $\hat{K}_1$ and $\hat{K}_2$ are bi-fuzzy languages over the events set $\hat{\Sigma}$. Then\par
  1) if $pr(\hat{K})$ = $\hat{K}$, then $pr(\hat{K}^{\uparrow})$ = $\hat{K}^{\uparrow}$;\par
  2) if $\hat{K}_1 \subseteq \hat{K}_2$, then $\hat{K}_{1}^{\uparrow} \subseteq \hat{K}_{2}^{\uparrow}$; \par
  3) $(\hat{K}_1 \cap \hat{K}_2 )^{\uparrow} \subseteq \hat{K}_{1}^{\uparrow} \cap \hat{K}_{2}^{\uparrow}$;\par
  4) $(\hat{K}_1 \cap \hat{K}_2 )^{\uparrow} = (\hat{K}_{1}^{\uparrow} \cap \hat{K}_{2}^{\uparrow})^{\uparrow}$;\par
  5) if $pr(\hat{K}_{1}^{\uparrow} \cap \hat{K}_{2}^{\uparrow}) = pr(\hat{K}_{1}^{\uparrow}) \cap pr(\hat{K}_{2}^{\uparrow}) $, then $(\hat{K}_{1} \cap \hat{K}_{2})^{\uparrow} = (\hat{K}_{1}^{\uparrow} \cap \hat{K}_{2}^{\uparrow})$;\par
  6) $\hat{K}_{1}^{\uparrow} \cup \hat{K}_{2}^{\uparrow} \subseteq (\hat{K}_1 \cup \hat{K}_2)^{\uparrow}$.
\end{proposition}
\begin{IEEEproof}
  1). Clearly, $\hat{K}^{\uparrow}$ is controllable, so is $pr(\hat{K}^{\uparrow})$. However, $pr(\hat{K}^{\uparrow}) \subseteq pr(\hat{K}) = \hat{K}$, which implies $pr(\hat{K}^{\uparrow}) \in C_{sub}(\hat{K})$. Hence, we have $pr(\hat{K}^{\uparrow}) \subseteq \hat{K}^{\uparrow}$, which together with the fact $ \hat{K}^{\uparrow} \subseteq pr(\hat{K}^{\uparrow})$, results in  $pr(\hat{K}^{\uparrow})$ = $\hat{K}^{\uparrow}$.   \par
  2). Since $\hat{K}_1 \subseteq \hat{K}_2$, for any $\hat{L} \in C_{sub}(\hat{K}_1)$, we have $\hat{L} \in C_{sub}(\hat{K}_2)$, i.e., $C_{sub}(\hat{K}_1) \subseteq C_{sub}(\hat{K}_2)$, which implies $\hat{K}_{1}^{\uparrow} \subseteq \hat{K}_{2}^{\uparrow}$ by 2) of Proposition 3.
  \par
  3). By means of  1) of Proposition 3, we have $(\hat{K}_{1} \cap \hat{K}_{2}) \subseteq \hat{K}_{1}$. From 2) of Proposition 7, we further have
  $(\hat{K}_{1} \cap \hat{K}_{2})^{\uparrow} \subseteq \hat{K}_{1}^{\uparrow}$. We can obtain $(\hat{K}_{1} \cap \hat{K}_{2})^{\uparrow} \subseteq \hat{K}_{2}^{\uparrow}$ similarly. Hence, by virtue of 6) of Proposition 3, we have $(\hat{K}_1 \cap \hat{K}_2 )^{\uparrow} \subseteq \hat{K}_{1}^{\uparrow} \cap \hat{K}_{2}^{\uparrow}$.
  \par
  4). From  3) of Proposition 7, we have $(\hat{K}_1 \cap \hat{K}_2 )^{\uparrow} \subseteq \hat{K}_{1}^{\uparrow} \cap \hat{K}_{2}^{\uparrow}$, together with the fact that $(\hat{K}_1 \cap \hat{K}_2 )^{\uparrow}$ is bi-fuzzy controllable, we obtain $(\hat{K}_1 \cap \hat{K}_2 )^{\uparrow} \subseteq (\hat{K}_{1}^{\uparrow} \cap \hat{K}_{2}^{\uparrow})^{\uparrow}$. The remainder is to show $ (\hat{K}_{1}^{\uparrow} \cap \hat{K}_{2}^{\uparrow})^{\uparrow}  \subseteq (\hat{K}_1 \cap \hat{K}_2 )^{\uparrow}$. By means of  3) of Proposition 3, the facts of $\hat{K}_{1}^{\uparrow} \subset \hat{K}_{1}$ and $\hat{K}_{2}^{\uparrow} \subset \hat{K}_{2}$ imply $ (\hat{K}_{1}^{\uparrow} \cap \hat{K}_{2}^{\uparrow})   \subseteq   (\hat{K}_1 \cap \hat{K}_2 ) $. From  2) of Proposition 7, we obtain $ (\hat{K}_{1}^{\uparrow} \cap \hat{K}_{2}^{\uparrow})^{\uparrow} \subseteq  (\hat{K}_1 \cap \hat{K}_2 )^{\uparrow} $. Therefore, we have $(\hat{K}_1 \cap \hat{K}_2 )^{\uparrow} = (\hat{K}_{1}^{\uparrow} \cap \hat{K}_{2}^{\uparrow})^{\uparrow}$.
  \par
  5). From  1) of Proposition 6 and 4) of Proposition 5, we obtain $\hat{K}_{1}^{\uparrow} \cap \hat{K}_{2}^{\uparrow}$ is bi-fuzzy controllable. Hence, we have $\hat{K}_{1}^{\uparrow} \cap \hat{K}_{2}^{\uparrow}= (\hat{K}_{1}^{\uparrow} \cap \hat{K}_{2}^{\uparrow})^{\uparrow}$, which together with the results from  4) of Proposition 7, results in $\hat{K}_{1}^{\uparrow} \cap \hat{K}_{2}^{\uparrow} = (\hat{K}_{1} \cap \hat{K}_{2})^{\uparrow}$.
  \par
  6). By virtue of  1) of Proposition 6 and  1) of Proposition 5, we obtain $\hat{K}_{1}^{\uparrow} \cup \hat{K}_{2}^{\uparrow}$ is bi-fuzzy controllable.
  By means of  4) of Proposition 3, the facts $\hat{K}_{1}^{\uparrow} \subset \hat{K}_{1}$ and $\hat{K}_{2}^{\uparrow} \subset \hat{K}_{2}$ imply
  $ \hat{K}_{1}^{\uparrow} \cup \hat{K}_{2}^{\uparrow} \subseteq (\hat{K}_1 \cup \hat{K}_2)$. Hence, we obtain $\hat{K}_{1}^{\uparrow} \cup \hat{K}_{2}^{\uparrow} \subseteq (\hat{K}_1 \cup \hat{K}_2)^{\uparrow}$.
\end{IEEEproof}

\begin{proposition}
  Suppose $\hat{K}$, $\hat{K}_1$ and $\hat{K}_2$ are bi-fuzzy languages over the events set $\hat{\Sigma}$. Then\par
  1) if $\hat{K}$ is controllable, then $\hat{K}^{\downarrow} = pr(\hat{K})$;\par
  2) if $\hat{K}_1 \subseteq \hat{K}_2$, then $\hat{K}_{1}^{\downarrow} \subseteq \hat{K}_{2}^{\downarrow}$; \par
  3) $(\hat{K}_1 \cap \hat{K}_2 )^{\downarrow} \subseteq (\hat{K}_{1}^{\downarrow} \cap \hat{K}_{2}^{\downarrow})^{\downarrow}$;\par
  4) $\hat{K}_{1}^{\downarrow} \cup \hat{K}_{2}^{\downarrow} = (\hat{K}_1 \cup \hat{K}_2)^{\downarrow}$.
\end{proposition}
\begin{IEEEproof}
  1). Since $\hat{K}$ is controllable, $pr(\hat{K})$ is also controllable. Together with the facts that $\hat{K} \subseteq pr(\hat{K})$ and $pr(\hat{K}) = pr(pr(\hat{K}))$, we obtain $pr(\hat{K}) \in C_{sup}(\hat{K})$. Hence, $\hat{K}^{\downarrow} \subseteq pr(\hat{K})$. Since for any $\hat{L} \in C_{sup}(\hat{K})$, we have $pr(\hat{K}) \subseteq pr(\hat{L}) = \hat{L}$, it further implies $pr(\hat{K}) \subseteq \bigcap_{\hat{L} \in C_{sup}(\hat{K})} \hat{L} = \hat{K}^{\downarrow}$ by means of  6) of Proposition 3. Therefore, we have $\hat{K}^{\downarrow} = pr(\hat{K})$.
  \par
  2). Since $\hat{K}_1 \subseteq \hat{K}_2$, for any $\hat{L} \in C_{sup}(\hat{K}_2)$, we have $\hat{L} \in C_{sup}(\hat{K}_1)$, i.e., $C_{sup}(\hat{K}_2) \subseteq C_{sup}(\hat{K}_1)$, it implies $\hat{K}_{1}^{\downarrow} \subseteq \hat{K}_{2}^{\downarrow}$ by virtue of  1) of Proposition 3.
  \par
  3). By virtue of  1) of Proposition 3, the facts that $\hat{K}_{1} \subseteq \hat{K}_{1}^{\downarrow}$ and $\hat{K}_{2} \subseteq \hat{K}_{2}^{\downarrow}$ imply $(\hat{K}_1 \cap \hat{K}_2 ) \subseteq (\hat{K}_{1}^{\downarrow} \cap \hat{K}_{2}^{\downarrow})$.
   According to  2) of Proposition 8, we obtain $(\hat{K}_1 \cap \hat{K}_2 )^{\downarrow} \subseteq (\hat{K}_{1}^{\downarrow} \cap \hat{K}_{2}^{\downarrow})^{\downarrow}$.
  \par
  4). By  2) of Proposition 3, we have $\hat{K}_{1} \subseteq (\hat{K}_1 \cup \hat{K}_2)$. By  2) of Proposition 8, we further have
  $\hat{K}_{1}^{\downarrow} \subseteq (\hat{K}_1 \cup \hat{K}_2)^{\downarrow}$. $\hat{K}_{2}^{\downarrow} \subseteq (\hat{K}_1 \cup \hat{K}_2)^{\downarrow}$ is obtained similarly. Then, by 5) of Proposition 3, we get $\hat{K}_{1}^{\downarrow} \cup \hat{K}_{2}^{\downarrow} \subseteq (\hat{K}_1 \cup \hat{K}_2)^{\downarrow}$. The remainder is to show $(\hat{K}_1 \cup \hat{K}_2)^{\downarrow} \subseteq   \hat{K}_{1}^{\downarrow} \cup \hat{K}_{2}^{\downarrow}$. Since the $\hat{K}_{1}^{\downarrow}$ and the $\hat{K}_{2}^{\downarrow}$ are controllable and prefix-close, by  1) and 2) of Proposition 5, $\hat{K}_{1}^{\downarrow} \cup \hat{K}_{2}^{\downarrow}$ is also
  controllable and prefix-close. Additionally,  by  4) of Proposition 3, the facts $\hat{K}_{1} \subset \hat{K}_{1}^{\downarrow}$ and $\hat{K}_{2} \subset \hat{K}_{2}^{\downarrow}$ imply that $(\hat{K}_1 \cup \hat{K}_2) \subseteq \hat{K}_{1}^{\downarrow} \cup \hat{K}_{2}^{\downarrow} $.
  Hence, we have $ \hat{K}_{1}^{\downarrow} \cup \hat{K}_{2}^{\downarrow} \in C_{sup}((\hat{K}_1 \cup \hat{K}_2))$. By means of  2) of Proposition 6, we obtain  $(\hat{K}_1 \cup \hat{K}_2)^{\downarrow} \subseteq   \hat{K}_{1}^{\downarrow} \cup \hat{K}_{2}^{\downarrow}$. Therefore, we complete the proof.
\end{IEEEproof}

\section{Proofs}
1. Proof of Proposition 1
    \begin{IEEEproof}
      We only prove the first and the third items, since the proofs for the others are similar. \\
      1) By virtue of commutative law and associative law, we have $\mu_1 \sqcap \mu_2 \sqcap \mu_1 = \mu_1 \sqcap \mu_1 \sqcap \mu_2$. According to idempotent law, we further obtain $\mu_1 \sqcap \mu_1 \sqcap \mu_2 = \mu_1 \sqcap \mu_2$. Therefore, we have $\mu_1 \sqcap \mu_2 \sqcap \mu_1 = \mu_1 \sqcap \mu_2$. With the definition of $\sqsubseteq$, we have $\mu_1 \sqcap \mu_2 \sqsubseteq \mu_1 $. \\
      3) According to commutative law and associative law, we have $(\mu_1 \sqcap \mu_3) \sqcap (\mu_2 \sqcap \mu_4) = (\mu_1 \sqcap \mu_2) \sqcap (\mu_3 \sqcap \mu_4)$. By means of the definition of $\sqsubseteq$, $\mu_1 \sqsubseteq \mu_2$ and $\mu_3 \sqsubseteq \mu_4 $ imply $\mu_1 \sqcap \mu_2 = \mu_1$ and $\mu_3 \sqcap \mu_4 = \mu_3$, respectively. Hence we have $(\mu_1 \sqcap \mu_3) \sqcap (\mu_2 \sqcap \mu_4) = \mu_1 \sqcap \mu_3$. With the definition of $\sqsubseteq$, we obtain $\mu_1 \sqcap \mu_3 \sqsubseteq  \mu_2 \sqcap \mu_4$.
    \end{IEEEproof}
\par
2. Proof of Proposition 2
 \begin{IEEEproof}
      We first show the left inclusion relation.
      Suppose that the system reaches the bi-fuzzy state $\hat{x}=\{\tilde{x}_1,\tilde{x}_2,\ldots,\tilde{x}_n\}$ after the occurrence of event $\hat{s}\hat{\sigma}$. Then with Equation (9), we have
        $L_{\hat{G}}(\hat{s}\hat{\sigma}) = \bigsqcup_{i\in [1,n]}[\tilde{x}_i \sqcap \frac{1}{1}] = \bigsqcup_{i\in [1,n]}\tilde{x}_i.$
         According to Equation (10), we have
         $
         L_{\hat{G}, m}(\hat{s}\hat{\sigma}) = \bigsqcup_{i\in [1,n]}[\tilde{x}_i \sqcap \tilde{x}_{m,i}].
         $
         By virtue of 1) of Proposition 1, we have $\tilde{x}_i \sqcap \tilde{x}_{m,i} \sqsubseteq \tilde{x}_i$. Furthermore, with  4) of Proposition 1, we obtain $\bigsqcup_{i\in [1,n]}[\tilde{x}_i \sqcap \tilde{x}_{m,i} ] \sqsubseteq \bigsqcup_{i\in [1,n]}\tilde{x}_i $. That is, $L_{\hat{G}, m}(\hat{s}\hat{\sigma})\sqsubseteq L_{\hat{G}}(\hat{s}\hat{\sigma})$.

      We continue to show the right inclusion relation.
      Suppose that the system reaches the bi-fuzzy state $\hat{x}=\{\tilde{x}_1,\tilde{x}_2,\ldots,\tilde{x}_n\}$ after the occurrence of event $\hat{s}$. Then we have
      \[
      L_{\hat{G}}(\hat{s}) = \bigsqcup_{i\in [1,n]}[\tilde{x}_i \sqcap \frac{1}{1}] = \bigsqcup_{i\in [1,n]}\tilde{x}_i.
      \]
      Assume the bi-fuzzy event $\hat{\sigma} = [\tilde{\sigma}_{i,j}]_{i,j \in [1,n]}$. By  commutative law, associative law and distributive law of \emph{NCFD}, we have
       \begin{align*}
       L_{\hat{G}}(\hat{s}\hat{\sigma})
       &= \bigsqcup_{j\in [1,n]}\{ \bigsqcup_{i \in [1,n]} [\tilde{x}_i \sqcap \tilde{\sigma}_{i,j}]\}    \\
       & = \bigsqcup_{i\in [1,n]}\{ \bigsqcup_{j \in [1,n]} [\tilde{x}_i \sqcap \tilde{\sigma}_{i,j}]\}   \\
       & = \bigsqcup_{i\in [1,n]}\{ \tilde{x}_i \sqcap [ \bigsqcup_{j \in [1,n]} \tilde{\sigma}_{i,j}]\}.
       \end{align*}
      According to 1) of Proposition 1, we have $\tilde{x}_i \sqcap [\bigsqcup_{j \in [1,n]} \tilde{\sigma}_{i,j}] \sqsubseteq \tilde{x}_i$ . By virtue of  4) of Proposition 1, we obtain $\bigsqcup_{i\in [1,n]}\{ \tilde{x}_i \sqcap [ \bigsqcup_{j \in [1,n]} \tilde{\sigma}_{i,j}]\} \sqsubseteq \bigsqcup_{i\in [1,n]}\tilde{x}_i $. That is,  $L_{\hat{G}}(\hat{s}\hat{\sigma})\sqsubseteq L_{\hat{G}}(\hat{s})$.
    \end{IEEEproof}
\par
3. Proof of Proposition 3
 \begin{IEEEproof}
      The proof of each item of the proposition relies on the corresponding item in Proposition 1. Therefore, we only prove the first item. The others are similar.\par
      1) For any $\hat{s} \in \hat{\Sigma}^{*}$, by  the definition of  $\cap$ and  1) of Proposition 1, we have $(\hat{L}_1 \cap \hat{L}_2)(\hat{s}) = (\hat{L}_1 (\hat{s}) \sqcap \hat{L}_2(\hat{s})) \sqsubseteq \hat{L}_1 (\hat{s}) $.
       With the definition of $\subseteq$, we obtain $\hat{L}_1 \cap \hat{L}_2 \subseteq \hat{L}_1 $.
    \end{IEEEproof}
\par

4. Proof of Proposition 4\par
The following Lemma is used to support the proof of  Proposition 4.
 \begin{lemma}
 Assume $\hat{A}$, $\hat{B}$, $\hat{C}$, and $\hat{D}$ are T2 fuzzy relation matrices for which  $\hat{A}\hat{\odot} \hat{C}$ and $\hat{B}\hat{\odot} \hat{D}$ are defined. Then
 $  (\hat{A} \tilde{\otimes} \hat{B})\hat{\odot}(\hat{C} \hat{\otimes} \hat{D}) = (\hat{A}\hat{\odot} \hat{C}) \hat{\otimes} (\hat{B}\hat{\odot} \hat{D}).$
\end{lemma}
    \begin{IEEEproof}
        Without loss of generality, suppose $\hat{A}, \hat{B}, \hat{C}, \hat{D}$ are $k*m, p*s, m*n, s*r$ matrices, respectively.
    Then we have\\
     \begin{align*}
      & (\hat{A} \hat{\otimes} \hat{B})\hat{\odot}(\hat{C} \hat{\otimes} \hat{D}) \\
      &=
           \left[ \begin{array}{ccc}
                \tilde{a}_{11}\sqcap \hat{B} & \ldots & \tilde{a}_{1m}\sqcap \hat{B} \\
                \vdots & \ddots & \vdots\\
                \tilde{a}_{k1}\sqcap \hat{B} & \ldots & \tilde{a}_{km}\sqcap \hat{B}
                            \end{array}
            \right]  \hat{\odot} \\
        &  \left[ \begin{array}{ccc}
                    \tilde{c}_{11}\sqcap \hat{D} & \ldots & \tilde{c}_{1n}\sqcap \hat{D} \\
                    \vdots & \ddots & \vdots\\
                    \tilde{c}_{m1}\sqcap \hat{D} & \ldots & \tilde{c}_{mn}\sqcap \hat{D}
                    \end{array}
          \right] \nonumber\\
        =& \bigg [\bigsqcup_{l=1}^{m}((\tilde{a}_{il}\sqcap \hat{B})\hat{\odot} (\tilde{c}_{lj}\sqcap \hat{D})) \bigg ]_{i\in[1, k]}^{j\in[1, n]} \nonumber\\
        =& \bigg [\bigsqcup_{l=1}^{m} \Big [\bigsqcup_{q=1}^{s}(\tilde{a}_{il}\sqcap \tilde{b}_{hq})\sqcap (\tilde{c}_{lj}\sqcap \tilde{d}_{qt}) \Big ]_{t \in [1,r]}^{h \in [1,p]} \bigg ]_{i\in[1, k]}^{j\in[1, n]}    \nonumber\\
        =& \bigg [\bigsqcup_{l=1}^{m} \Big [\bigsqcup_{q=1}^{s} (\tilde{a}_{il}\sqcap \tilde{c}_{lj})\sqcap (\tilde{b}_{hq}\sqcap \tilde{d}_{qt}) \Big ]_{t \in [1,r]}^{h \in [1,p]} \bigg ]_{i\in[1, k]}^{j\in[1, n]} \ \   \nonumber\\
        =& \bigg [\bigsqcup_{l=1}^{m} (\tilde{a}_{il}\sqcap \tilde{c}_{lj})\sqcap  \Big [\bigsqcup_{q=1}^{s} (\tilde{b}_{hq}\sqcap \tilde{d}_{qt}) \Big ]_{t \in [1,r]}^{h \in [1,p]} \bigg ]_{i\in[1, k]}^{j\in[1, n]}\ \     \nonumber\\
        =& \bigg[ \Big [\hat{A} \hat{\odot} \hat{C} \Big ]_{ij} \sqcap (\hat{B}\hat{\odot} \hat{D})\bigg ]_{i\in[1, k]}^{j\in[1, n]}  \nonumber\\
        =& (\hat{A}\hat{\odot} \hat{C})\hat{\otimes} (\hat{B}\hat{\odot} \hat{D}). \nonumber
    \end{align*}
 \end{IEEEproof}
The proof of Proposition 4:\par
\begin{IEEEproof}
 Let $|X_{1}|=m$ and $|X_{2}|=n$. Suppose for any $\hat{s}\in\hat{\Sigma}^{*}$ with $\hat{s} = \hat{\sigma}^{1}\hat{\sigma}^{2}\ldots\hat{\sigma}^{k}$, the corresponding matrices of bi-fuzzy event $\hat{\sigma}^{i}, i\in[1, k]$ in $\hat{G}_{1}$ and $\hat{G}_{2}$  are denoted by $\hat{\sigma}^{i}_{1}$ and $\hat{\sigma}^{i}_{2}$, respectively. For convenience, let $ (\hat{\sigma}^{1}_{1}\hat{\odot}\hat{\sigma}^{2}_{1} \ldots\hat{\sigma}^{k}_{1}) =  \hat{\sigma}^{s}_{1}$ and $(\hat{\sigma}^{1}_{2}\hat{\odot}\hat{\sigma}^{2}_{2}\ldots\hat{\sigma}^{k}_{2}) = \hat{\sigma}^{s}_{2} $. Suppose that the bi-fuzzy automata $\hat{G}_{1}$ and $\hat{G}_{2}$ reach the bi-fuzzy states $\hat{x}_1 = \{\tilde{x}_{11},\tilde{x}_{12},\ldots \tilde{x}_{1m}\}$ and $\hat{x}_2 = \{\tilde{x}_{21},\tilde{x}_{22},\ldots \tilde{x}_{2n}\}$, respectively, after the occurrence of events string $\hat{s}$. Then we have:
\begin{align}
L(\hat{G}_{1}||\hat{G}_{2})(\hat{s}) &= (\hat{x}_{01}\hat{\otimes} \hat{x}_{02})\hat{\odot} (\hat{\sigma}^{s}_{1} \hat{\otimes} \hat{\sigma}^{s}_{2})\hat{\odot} \hat{A}^{T}_{m*n} \nonumber\\
&=(\hat{x}_{01} \hat{\odot} \hat{\sigma}^{s}_{1})  \hat{\otimes}( \hat{x}_{02} \hat{\odot}  \hat{\sigma}^{s}_{2}) \hat{\odot} \hat{A}^{T}_{m*n}  \nonumber\\
&=\bigsqcup_{i \in [1,m ]}\bigsqcup_{j \in [1,n]} [ \hat{x}_{1i} \sqcap \hat{x}_{2j} ] \nonumber\\
&=\bigsqcup_{i \in [1,m ]} [ \hat{x}_{1i} \sqcap   \bigsqcup_{j \in [1,n]}  \hat{x}_{2j}  ]   \nonumber\\
&= ( \bigsqcup_{i \in [1,m ]} \hat{x}_{1i} ) \sqcap  ( \bigsqcup_{j \in [1,n]}  \hat{x}_{2j})    \nonumber\\
&=L(\hat{G}_{1})(\hat{s}) \sqcap L(\hat{G}_{2})(\hat{s}). \nonumber
\end{align}
\end{IEEEproof}
\par

\section*{Acknowledgments}

This work is supported in part by the National
Natural Science Foundation (Nos. 61272058, 61073054), the Natural
Science Foundation of Guangdong Province of China (No.
10251027501000004),  the Specialized Research Fund for the Doctoral Program of Higher Education of China
(No. 20100171110042).

\ifCLASSOPTIONcaptionsoff
  \newpage
\fi

\nocite{*}

\begin{IEEEbiography}[{\includegraphics[width=1in,height=1.25in,clip,keepaspectratio]{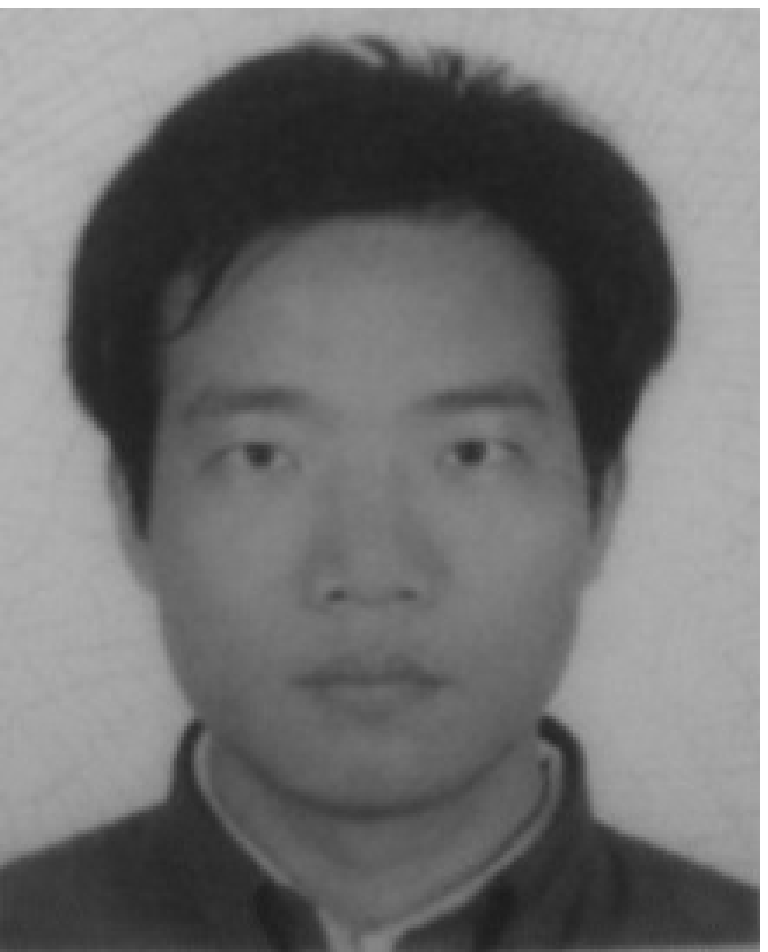}}]
{Weilin Deng}
 received the B.S.  degree and the M.S. degree in computer science from the South China University of Technology, Guangzhou, China, in 2003 and 2008, respectively. Since 2011, he has been working toward the Ph.D. degree in the Department of Computer Science, Sun Yat-sen University,
 Guangzhou, China.
 \par
His main research interests include fuzzy discrete event systems, supervisory control, and decentralized control.
\end{IEEEbiography}

\begin{IEEEbiography}[{\includegraphics[width=1.0in,height=1.25in,clip,keepaspectratio]{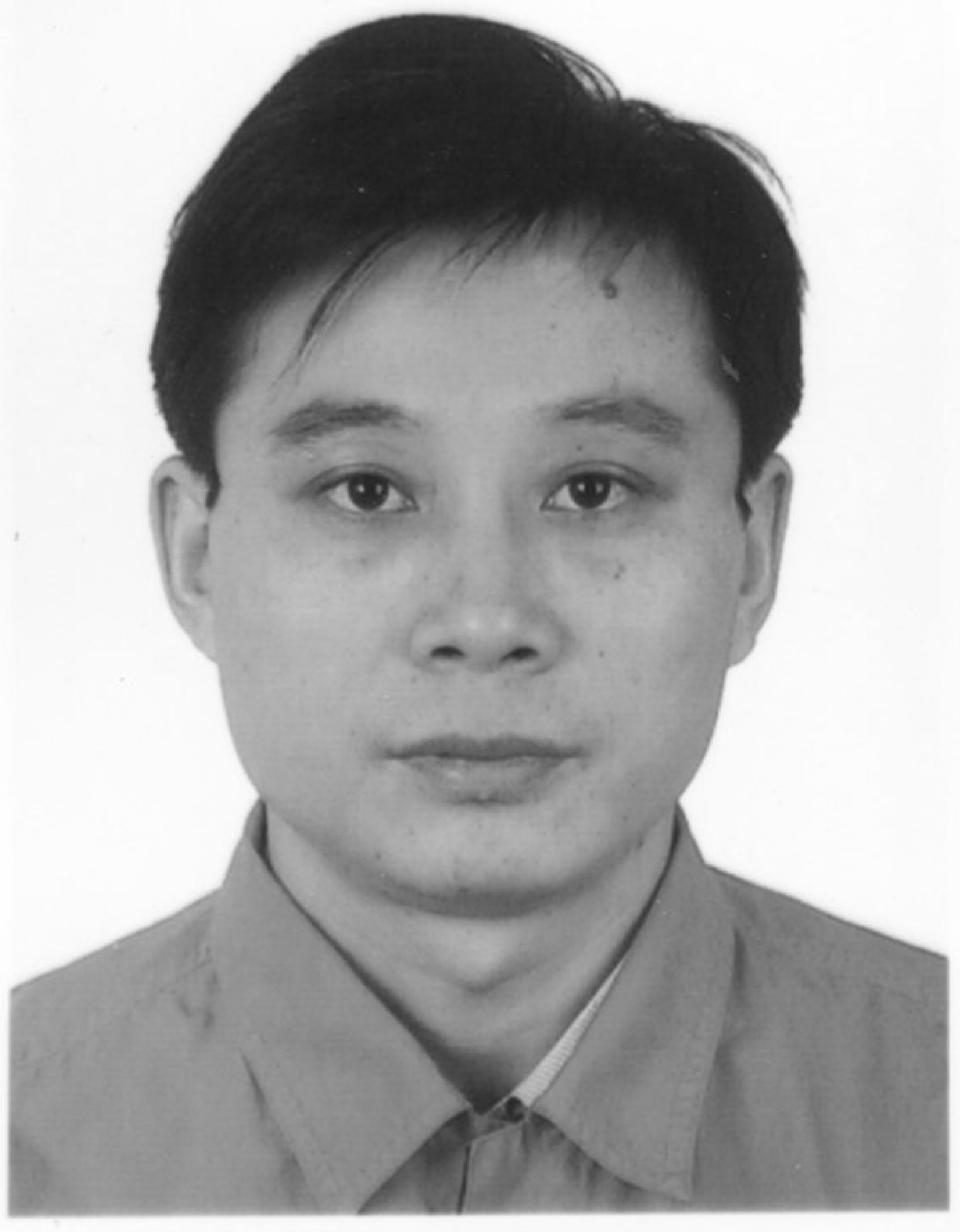}}]
{Daowen Qiu}

received the M.S. degree in mathematics in 1993 from Jiangxi Normal University, Nachang, China, and then he received the Ph.D. degree in mathematics from Sun Yat-Sen University, Guangzhou, China, in 2000. He completed the postdoctoral research in computer science at Tsinghua University, Beijing, China, in 2002.

Since 2004, he has been a Professor of computer science at Sun Yat-Sen University. He is interested in models of nonclassical computation (including quantum, fuzzy and probabilistic computation) and quantum information theory. He has published over 120 peer-review papers in academic journals and international conferences, including: Information and Computation, Artificial Intelligence, Journal of Computer and System Sciences, Theoretical Computer Science, IEEE Transactions on Automatic Control, IEEE Transactions on SMC-Part B, IEEE Transactions on Fuzzy Systems, Physical Review A, Quantum Information and Computation, Journal of Physics A, Science in China.

\end{IEEEbiography}

%
%


\end{document}